\documentclass[11pt,english]{article}
\usepackage{lmodern}

\usepackage[T1]{fontenc}
\usepackage[latin9]{inputenc}
\usepackage{geometry}
\usepackage{fancybox}
\usepackage{calc}
\usepackage{units}
\usepackage{booktabs}
\usepackage{dsfont}
\usepackage{amsmath}
\usepackage{amsthm}
\usepackage{amssymb}
\usepackage{csquotes}
\allowdisplaybreaks

\makeatletter
\numberwithin{figure}{section}
\numberwithin{equation}{section}
\theoremstyle{plain}
\newtheorem{theorem}{\protect\theoremname}
\newtheorem*{theorem*}{\protect\theoremname}
\theoremstyle{plain}
\newtheorem{lemma}[theorem]{\protect\lemmaname}
\theoremstyle{plain}
\newtheorem*{lemma*}{\protect\lemmaname}
\theoremstyle{plain}
\newtheorem{proposition}{Proposition}
\theoremstyle{definition}
\newtheorem{remark}{Remark}
\theoremstyle{definition}
\newtheorem{definition}[theorem]{Definition}
\theoremstyle{plain}
\newtheorem{corollary}[theorem]{Corollary}
\theoremstyle{plain}
\newtheorem*{corollary*}{Corollary}
\usepackage{comment,url,algorithm,algorithmic,graphicx,subcaption,relsize}
\usepackage{amssymb,amsfonts,amsmath,amsthm,amscd,dsfont,mathrsfs,mathtools,microtype,nicefrac,pifont}
\usepackage{float,psfrag,epsfig,color,url,hyperref}
\hypersetup{
  colorlinks,
  linkcolor={red!50!black},
  citecolor={blue!50!black},
  urlcolor={blue!80!black}
}
\usepackage{upgreek}
\usepackage[dvipsnames]{xcolor}
\usepackage{epstopdf,bbm,mathtools,enumitem}
\usepackage[toc,page]{appendix}
\usepackage{dsfont}

\usepackage[mathscr]{euscript}
\usepackage{custom}

\makeatletter
\renewcommand{\paragraph}{%
  \@startsection{paragraph}{4}%
  {\z@}{1.25ex \@plus 1ex \@minus .2ex}{-1em}%
  {\normalfont\normalsize\bfseries}%
}
\makeatother

\makeatother

\usepackage{babel}
\providecommand{\lemmaname}{Lemma}
\providecommand{\theoremname}{Theorem}

\begin{document}
\def\balign#1\ealign{\begin{align}#1\end{align}}
\def\baligns#1\ealigns{\begin{align*}#1\end{align*}}
\def\balignat#1\ealign{\begin{alignat}#1\end{alignat}}
\def\balignats#1\ealigns{\begin{alignat*}#1\end{alignat*}}
\def\bitemize#1\eitemize{\begin{itemize}#1\end{itemize}}
\def\benumerate#1\eenumerate{\begin{enumerate}#1\end{enumerate}}

\newenvironment{talign*}
 {\let\displaystyle\textstyle\csname align*\endcsname}
 {\endalign}
\newenvironment{talign}
 {\let\displaystyle\textstyle\csname align\endcsname}
 {\endalign}

\def\balignst#1\ealignst{\begin{talign*}#1\end{talign*}}
\def\balignt#1\ealignt{\begin{talign}#1\end{talign}}

\let\originalleft\left
\let\originalright\right
\renewcommand{\left}{\mathopen{}\mathclose\bgroup\originalleft}
\renewcommand{\right}{\aftergroup\egroup\originalright}

\def\Gronwall{Gr\"onwall\xspace}
\def\Holder{H\"older\xspace}
\def\Ito{It\^o\xspace}
\def\Nystrom{Nystr\"om\xspace}
\def\Schatten{Sch\"atten\xspace}
\def\Matern{Mat\'ern\xspace}

\def\tinycitep*#1{{\tiny\citep*{#1}}}
\def\tinycitealt*#1{{\tiny\citealt*{#1}}}
\def\tinycite*#1{{\tiny\cite*{#1}}}
\def\smallcitep*#1{{\scriptsize\citep*{#1}}}
\def\smallcitealt*#1{{\scriptsize\citealt*{#1}}}
\def\smallcite*#1{{\scriptsize\cite*{#1}}}

\def\blue#1{\textcolor{blue}{{#1}}}
\def\green#1{\textcolor{green}{{#1}}}
\def\orange#1{\textcolor{orange}{{#1}}}
\def\purple#1{\textcolor{purple}{{#1}}}
\def\red#1{\textcolor{red}{{#1}}}
\def\teal#1{\textcolor{teal}{{#1}}}

\def\mbi#1{\boldsymbol{#1}} 
\def\mbf#1{\mathbf{#1}}
\def\mrm#1{\mathrm{#1}}
\def\tbf#1{\textbf{#1}}
\def\tsc#1{\textsc{#1}}

\def\mbiA{\mbi{A}}
\def\mbiB{\mbi{B}}
\def\mbiC{\mbi{C}}
\def\mbiDelta{\mbi{\Delta}}
\def\mbif{\mbi{f}}
\def\mbiF{\mbi{F}}
\def\mbih{\mbi{g}}
\def\mbiG{\mbi{G}}
\def\mbih{\mbi{h}}
\def\mbiH{\mbi{H}}
\def\mbiI{\mbi{I}}
\def\mbim{\mbi{m}}
\def\mbiP{\mbi{P}}
\def\mbiQ{\mbi{Q}}
\def\mbiR{\mbi{R}}
\def\mbiv{\mbi{v}}
\def\mbiV{\mbi{V}}
\def\mbiW{\mbi{W}}
\def\mbiX{\mbi{X}}
\def\mbiY{\mbi{Y}}
\def\mbiZ{\mbi{Z}}

\def\textsum{{\textstyle\sum}} 
\def\textprod{{\textstyle\prod}} 
\def\textbigcap{{\textstyle\bigcap}} 
\def\textbigcup{{\textstyle\bigcup}} 

\def\reals{\mathbb{R}} 
\def\integers{\mathbb{Z}} 
\def\rationals{\mathbb{Q}} 
\def\naturals{\mathbb{N}} 
\def\complex{\mathbb{C}} 

\def\what#1{\widehat{#1}}

\def\twovec#1#2{\left[\begin{array}{c}{#1} \\ {#2}\end{array}\right]}
\def\threevec#1#2#3{\left[\begin{array}{c}{#1} \\ {#2} \\ {#3} \end{array}\right]}
\def\nvec#1#2#3{\left[\begin{array}{c}{#1} \\ {#2} \\ \vdots \\ {#3}\end{array}\right]} 

\def\maxeig#1{\lambda_{\mathrm{max}}\left({#1}\right)}
\def\mineig#1{\lambda_{\mathrm{min}}\left({#1}\right)}

\def\Re{\operatorname{Re}} 
\def\indic#1{\mbb{I}\left[{#1}\right]} 
\def\logarg#1{\log\left({#1}\right)} 
\def\polylog{\operatorname{polylog}}
\def\maxarg#1{\max\left({#1}\right)} 
\def\minarg#1{\min\left({#1}\right)} 
\def\Earg#1{\E\left[{#1}\right]}
\def\Esub#1{\E_{#1}}
\def\Esubarg#1#2{\E_{#1}\left[{#2}\right]}
\def\bigO#1{\mathcal{O}\left(#1\right)} 
\def\littleO#1{o(#1)} 
\def\P{\mbb{P}} 
\def\Parg#1{\P\left({#1}\right)}
\def\Psubarg#1#2{\P_{#1}\left[{#2}\right]}
\def\Trarg#1{\Tr\left[{#1}\right]} 
\def\trarg#1{\tr\left[{#1}\right]} 
\def\Var{\mrm{Var}} 
\def\Vararg#1{\Var\left[{#1}\right]}
\def\Varsubarg#1#2{\Var_{#1}\left[{#2}\right]}
\def\Cov{\operatorname{Cov}} 
\def\Covarg#1{\Cov\left[{#1}\right]}
\def\Covsubarg#1#2{\Cov_{#1}\left[{#2}\right]}
\def\Corr{\mrm{Corr}} 
\def\Corrarg#1{\Corr\left[{#1}\right]}
\def\Corrsubarg#1#2{\Corr_{#1}\left[{#2}\right]}
\newcommand{\info}[3][{}]{\mathbb{I}_{#1}\left({#2};{#3}\right)} 
\newcommand{\staticexp}[1]{\operatorname{exp}(#1)} 
\newcommand{\para}[1]{\medskip\noindent\textbf{#1}\,} %
\newcommand{\loglihood}[0]{\mathcal{L}} 

\providecommand{\arccos}{\mathop\mathrm{arccos}}
\providecommand{\dom}{\mathop\mathrm{dom}}
\providecommand{\diag}{\mathop\mathrm{diag}}
\providecommand{\tr}{\mathop\mathrm{tr}}
\providecommand{\card}{\mathop\mathrm{card}}
\providecommand{\sign}{\mathop\mathrm{sign}}
\providecommand{\conv}{\mathop\mathrm{conv}} 
\def\rank#1{\mathrm{rank}({#1})}
\def\supp#1{\mathrm{supp}({#1})}

\providecommand{\minimize}{\mathop\mathrm{minimize}}
\providecommand{\maximize}{\mathop\mathrm{maximize}}
\providecommand{\subjectto}{\mathop\mathrm{subject\;to}}

\def\openright#1#2{\left[{#1}, {#2}\right)}

\ifdefined\nonewproofenvironments\else
\ifdefined\ispres\else
\newenvironment{proof-of-lemma}[1][{}]{\noindent\textbf{Proof of Lemma {#1}}
\hspace*{1em}}{\qed\\}
 \newenvironment{proof-of-theorem}[1][{}]{\noindent\textbf{Proof of Theorem {#1}}
   \hspace*{1em}}{\qed\\}

\newtheorem*{rem*}{Remark}
\newenvironment{rem}{\noindent\textbf{Remark.}
  \hspace*{0em}}{\smallskip}
\newenvironment{remarks}{\noindent\textbf{Remarks}
  \hspace*{1em}}{\smallskip}
\fi
\fi
\makeatletter
\@addtoreset{equation}{section}
\makeatother
\def\theequation{\thesection.\arabic{equation}}

\newcommand{\cmark}{\ding{51}}

\newcommand{\xmark}{\ding{55}}

\newcommand{\eq}[1]{\begin{align}#1\end{align}}
\newcommand{\eqn}[1]{\begin{align*}#1\end{align*}}
\renewcommand{\Pr}[1]{\mathbb{P}\left( #1 \right)}
\newcommand{\Ex}[1]{\mathbb{E}\left[#1\right]}

\newcommand{\matt}[1]{{\textcolor{Maroon}{[Matt: #1]}}}
\newcommand{\kook}[1]{{\textcolor{blue}{[Kook: #1]}}}
\definecolor{OliveGreen}{rgb}{0,0.6,0}
\newcommand{\sv}[1]{{\textcolor{OliveGreen}{[Santosh: #1]}}}

\newcommand{\tv}{\mathsf{TV}}

\global\long\def\on#1{\operatorname{#1}}%

\global\long\def\bw{\mathsf{Ball\ walk}}%
\global\long\def\sw{\mathsf{Speedy\ walk}}%
\global\long\def\gw{\mathsf{Gaussian\ walk}}%
\global\long\def\dw{\mathsf{Dikin\ walk}}%
\global\long\def\rhmc{\mathsf{Riemannian\ Hamiltonian\ Monte\ Carlo}}%
\global\long\def\ml{\mathsf{Mirror\ Langevin}}%
\global\long\def\fb{\mathsf{In\text{-}and\text{-}Out}}%
\global\long\def\har{\mathsf{Hit\text{-}and\text{-}Run}}%

\global\long\def\O{O}%
\global\long\def\Otilde{\widetilde{O}}%

\global\long\def\E{\mathbb{E}}%
\global\long\def\Z{\mathbb{Z}}%
\global\long\def\P{\mathbb{P}}%
\global\long\def\N{\mathbb{N}}%

\global\long\def\R{\mathbb{R}}%
\global\long\def\Rd{\mathbb{R}^{d}}%
\global\long\def\Rdd{\mathbb{R}^{d\times d}}%
\global\long\def\Rn{\mathbb{R}^{n}}%
\global\long\def\Rnn{\mathbb{R}^{n\times n}}%

\global\long\def\psd{\mathbb{S}_{+}^{d}}%
\global\long\def\pd{\mathbb{S}_{++}^{d}}%

\global\long\def\defeq{\stackrel{\mathrm{{\scriptscriptstyle def}}}{=}}%

\global\long\def\veps{\varepsilon}%
\global\long\def\lda{\lambda}%
\global\long\def\vphi{\varphi}%

\global\long\def\half{\frac{1}{2}}%
\global\long\def\nhalf{\nicefrac{1}{2}}%
\global\long\def\texthalf{{\textstyle \frac{1}{2}}}%

\global\long\def\ind{\mathds{1}}%

\global\long\def\chooses#1#2{_{#1}C_{#2}}%

\global\long\def\vol{\on{vol}}%

\global\long\def\law{\msf{law}}%

\global\long\def\tr{\on{tr}}%

\global\long\def\diag{\on{diag}}%

\global\long\def\Diag{\on{Diag}}%

\global\long\def\inter{\on{int}}%

\global\long\def\e{\mathrm{e}}%

\global\long\def\id{\mathrm{id}}%

\global\long\def\spanning{\on{span}}%

\global\long\def\rows{\on{row}}%

\global\long\def\cols{\on{col}}%

\global\long\def\rank{\on{rank}}%

\global\long\def\T{\mathsf{T}}%

\global\long\def\bs#1{\boldsymbol{#1}}%

\global\long\def\eu#1{\EuScript{#1}}%

\global\long\def\mb#1{\mathbf{#1}}%

\global\long\def\mbb#1{\mathbb{#1}}%

\global\long\def\mc#1{\mathcal{#1}}%

\global\long\def\mf#1{\mathfrak{#1}}%

\global\long\def\ms#1{\mathscr{#1}}%

\global\long\def\mss#1{\mathsf{#1}}%

\global\long\def\msf#1{\mathsf{#1}}%

\global\long\def\textint{{\textstyle \int}}%
\global\long\def\Dd{\mathrm{D}}%
\global\long\def\D{\mathrm{d}}%
\global\long\def\grad{\nabla}%
 
\global\long\def\hess{\nabla^{2}}%
 
\global\long\def\lapl{\triangle}%
 
\global\long\def\deriv#1#2{\frac{\D#1}{\D#2}}%
 
\global\long\def\pderiv#1#2{\frac{\partial#1}{\partial#2}}%
 
\global\long\def\de{\partial}%
\global\long\def\lagrange{\mathcal{L}}%
\global\long\def\Div{\on{div}}%

\global\long\def\Gsn{\mathcal{N}}%
 
\global\long\def\BeP{\textnormal{BeP}}%
 
\global\long\def\Ber{\textnormal{Ber}}%
 
\global\long\def\Bern{\textnormal{Bern}}%
 
\global\long\def\Bet{\textnormal{Beta}}%
 
\global\long\def\Beta{\textnormal{Beta}}%
 
\global\long\def\Bin{\textnormal{Bin}}%
 
\global\long\def\BP{\textnormal{BP}}%
 
\global\long\def\Dir{\textnormal{Dir}}%
 
\global\long\def\DP{\textnormal{DP}}%
 
\global\long\def\Expo{\textnormal{Expo}}%
 
\global\long\def\Gam{\textnormal{Gamma}}%
 
\global\long\def\GEM{\textnormal{GEM}}%
 
\global\long\def\HypGeo{\textnormal{HypGeo}}%
 
\global\long\def\Mult{\textnormal{Mult}}%
 
\global\long\def\NegMult{\textnormal{NegMult}}%
 
\global\long\def\Poi{\textnormal{Poi}}%
 
\global\long\def\Pois{\textnormal{Pois}}%
 
\global\long\def\Unif{\textnormal{Unif}}%

\global\long\def\bpar#1{\bigl(#1\bigr)}%
\global\long\def\Bpar#1{\Bigl(#1\Bigr)}%

\global\long\def\snorm#1{\|#1\|}%
\global\long\def\bnorm#1{\bigl\Vert#1\bigr\Vert}%
\global\long\def\Bnorm#1{\Bigl\Vert#1\Bigr\Vert}%

\global\long\def\sbrack#1{[#1]}%
\global\long\def\bbrack#1{\bigl[#1\bigr]}%
\global\long\def\Bbrack#1{\Bigl[#1\Bigr]}%

\global\long\def\sbrace#1{\{#1\}}%
\global\long\def\bbrace#1{\bigl\{#1\bigr\}}%
\global\long\def\Bbrace#1{\Bigl\{#1\Bigr\}}%

\global\long\def\Abs#1{\left\lvert #1\right\rvert }%
\global\long\def\Par#1{\left(#1\right)}%
\global\long\def\Brack#1{\left[#1\right]}%
\global\long\def\Brace#1{\left\{  #1\right\}  }%

\global\long\def\inner#1{\langle#1\rangle}%
\global\long\def\binner#1{\bigl\langle#1\bigr\rangle}%
\global\long\def\Binner#1{\Bigl\langle#1\Bigr\rangle}%

\global\long\def\norm#1{\lVert#1\rVert}%
\global\long\def\onenorm#1{\norm{#1}_{1}}%
\global\long\def\twonorm#1{\norm{#1}_{2}}%
\global\long\def\infnorm#1{\norm{#1}_{\infty}}%
\global\long\def\fronorm#1{\norm{#1}_{\text{F}}}%
\global\long\def\nucnorm#1{\norm{#1}_{*}}%
\global\long\def\staticnorm#1{\|#1\|}%
\global\long\def\statictwonorm#1{\staticnorm{#1}_{2}}%

\global\long\def\mmid{\mathbin{\|}}%

\global\long\def\otilde#1{\widetilde{\mc O}(#1)}%
\global\long\def\wtilde{\widetilde{W}}%
\global\long\def\wt#1{\widetilde{#1}}%

\global\long\def\KL{\msf{KL}}%
\global\long\def\FI{\msf{FI}}%
\global\long\def\Ent{\msf{Ent}}%
\global\long\def\dtv{d_{\mathsf{TV}}}%

\global\long\def\cov{\operatorname{cov}}%
\global\long\def\var{\operatorname{var}}%

\global\long\def\cred#1{\textcolor{red}{#1}}%
\global\long\def\cblue#1{\textcolor{blue}{#1}}%
\global\long\def\cgreen#1{\textcolor{green}{#1}}%
\global\long\def\ccyan#1{\textcolor{cyan}{#1}}%

\global\long\def\iff{\Leftrightarrow}%
 
\global\long\def\textfrac#1#2{{\textstyle \frac{#1}{#2}}}%

\newcommand{\Pac}{\mc P_{2,\rm ac}(\R^d)}
\title{In-and-Out: Algorithmic Diffusion for Sampling Convex Bodies}

\author{\!\!\!
Yunbum Kook\thanks{
  School of Computer Science,
  Georgia Institute of Technology, \texttt{yb.kook@gatech.edu}
} \ \ \ \ \ \ \ \
 Santosh S. Vempala\thanks{
  School of Computer Science,
  Georgia Institute of Technology, \texttt{vempala@gatech.edu}
 }\ \ \ \ \ \ \ \
Matthew S.\ Zhang\thanks{
  Department of Computer Science,
  University of Toronto, and Vector Institute, \texttt{matthew.zhang@mail.utoronto.ca}
}
}
\maketitle

\begin{abstract}
    We present a new random walk for uniformly sampling high-dimensional convex bodies. It achieves state-of-the-art runtime complexity with stronger guarantees on the output than previously known, namely in R\'enyi divergence (which implies TV, $\mathcal{W}_2$, KL, $\chi^2$). The proof departs from known approaches for polytime algorithms for the problem --- we utilize a stochastic diffusion perspective to show contraction to the target distribution with the rate of convergence determined by functional isoperimetric constants of the target distribution.
\end{abstract}

\section{Introduction}
Generating random samples from a high-dimensional convex body is a basic algorithmic problem with myriad connections and applications. The core of the celebrated result of~\cite{dyer1991random}, giving a randomized polynomial-time algorithm for computing the volume of a convex body, was the first polynomial-time algorithm for uniformly sampling convex bodies. 
In the decades since, the study of sampling has led to a long series of improvements in its algorithmic complexity~\cite{lovasz1990mixing,lovasz1993random,kannan1997random,lovasz2006hit,cousins2018gaussian}, often based on uncovering new mathematical/geometric structure, establishing connections to other fields (e.g., functional analysis, matrix concentration) and developing new tools for proving isoperimetric inequalities and analyzing Markov chains. With the proliferation of data and the increasing importance of machine learning, sampling has also become an essential algorithmic tool, with applications needing samplers in very high dimension, e.g.,
scientific computing~\cite{cousins2016practical,haraldsdottir2017chrr,kook2022sampling}, systems biology~\cite{lewis2012constraining, thiele2013community}, differential privacy~\cite{mcsherry2007mechanism,mironov2017renyi} and machine learning~\cite{bingham2019pyro,stan}.

Samplers for convex bodies are based on Markov chains (see Sec. \S\ref{sec:related} for a summary). Their analysis is based on bounding the {\em conductance} of the associated Markov chain, which in turn bounds the mixing rate. Analyzing the conductance requires combining delicate geometric arguments with (Cheeger) isoperimetric inequalities for convex bodies. An archetypal example of the latter is the following:
for any measurable partition $S_1, S_2, S_3$ of a convex body $\mc K \subset \Rd$, we have
\[
\vol(S_3) \ge \frac{d(S_1, S_2)}{C_{\mc K}} \min \{ \vol(S_1), \vol(S_2)\}\,,
\]
where $d(\cdot, \cdot)$ is the (minimum) Euclidean distance, and $C_{\mc K}$ is an isoperimetric constant of the uniform distribution over $\mc K$. (The KLS conjecture posits that $C_{\mc K} = \O(1)$ for any convex body $\mc K$ in {\em isotropic position}, i.e., under the normalization that a random point from $\mc K$ has identity covariance). The coefficient $C_{\mc K}^2$ is bounded by the Poincar\'e constant of the uniform distribution over $\mc K$ (and they are in fact asymptotically equal). The classical proof of conductance uses geometric properties of the random walk at hand to reduce the analysis to a suitable isoperimetric inequality (see e.g., \cite{lovasz1993random,vempala2005geometric}). The end result is a guarantee on the number of steps after which the total variation distance (TV distance) between the current distribution and the target is bounded by a desired error parameter. 
This framework has been widely used and effective in analyzing an array of candidate samplers, e.g., $\bw$ \cite{kannan1997random}, $\har$ \cite{lovasz1999hit,lovasz2006hit}, $\rhmc$ \cite{lee2018convergence} etc.

A different approach, studied intensively over the past decade, is based on {\em diffusion}. The basic idea is to first analyze a continuous-time diffusion process, typically modeled by a \emph{stochastic differential equation} (SDE), and then show that a suitable time-discretization of the process, sometimes together with a Metropolis filter, converges to the desired distribution efficiently. 
A major success along this line is the $\msf{Unadjusted\ Langevin\ Algorithm}$ and its variants, studied first in \cite{besag1995bayesian}. These algorithms have strong guarantees for sampling ``nice'' distributions~\cite{dalalyan2012sparse,dalalyan2017further,durmus2019analysis,vempala2019rapid}, such as ones that are strongly logconcave, or more generally distributions satisfying isoperimetric inequalities, while also obeying some smoothness conditions. The analysis of these algorithms is markedly different from the conductance approach, and typically yields guarantees in stronger metrics such as the $\KL$-divergence. 

Our starting point is the following question: 
\begin{center}
{\em Can diffusion-based approaches be used for the problem of sampling convex bodies?}
\end{center}
Despite remarkable progress, thus far, constrained sampling problems have evaded the diffusion approach, except as a high-level analogy (e.g., the $\bw$ can be viewed as a discretization of Brownian motion, but this alone does not suggest a route for analysis) or with significantly worse convergence rates (e.g., \cite{pmlr-v65-brosse17a, bubeck2018sampling}). 

Our main finding is a simple diffusion-based algorithm that can be mapped to a stochastic process such that the rate of convergence is bounded directly by an appropriate functional inequality for the target distribution. As a consequence, for the first time, we obtain clean end-to-end guarantees in the R\'enyi divergence (which implies guarantees in other well-known quantities such as $\mc W_2, \tv, \KL, \chi^2$ etc.), while giving state-of-the-art runtime complexity for sampling convex bodies (e.g., $\bw$ or $\sw$ \cite{lovasz1993random,kannan1997random}).
Besides being a stronger guarantee on the output, R\'enyi divergence is of particular interest for differential privacy~\cite{mironov2017renyi}. Perhaps most interesting is that our proof approach is quite different from prior work on convex body sampling.  In summary,
\begin{itemize}
    \item The guarantees hold for the $q$-R\'enyi divergences while matching the rates of previous work (prior work only had guarantees in the TV distance). 
    \item The analysis is simple, modular, and easily extendable to several other settings.
\end{itemize}

\begin{figure}[t]
\includegraphics[width=0.8\textwidth]{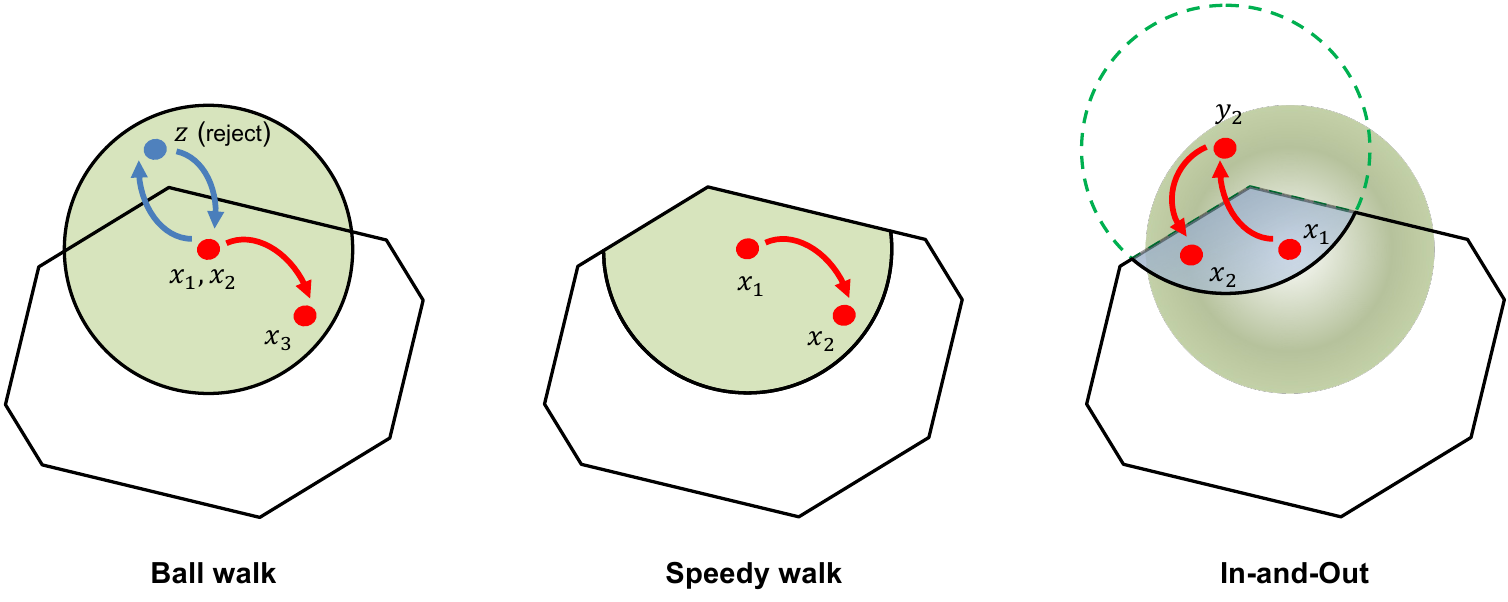}
\centering
\caption{Description of uniform samplers: (i) $\bw$: proposes a uniform random point $z$ from $B_{\delta}(x_1)$, but $z\notin \mc K$ so it stays at $x_1=x_2$. (ii) $\sw$: moves to $x_2$ drawn uniformly at random from $\mc K \cap B_{\delta}(x_1)$. (iii) $\fb$: first moves to $y_2$ obtained by taking a Gaussian step from $x_1$, and then to $x_2$ obtained by sampling the truncated Gaussian $\mc N(y_2,hI_d)|_{\mc K}$.}\label{fig:description}
\end{figure}

\subsection{Diffusion for uniform sampling}\label{sec:alg-stoch}

We propose the following $\fb$\footnote{This name reflects the ``geometry'' of how the iterates are moving. As we elaborate in Remark~\ref{rmk:proximal}, the name `proximal sampler' may be more familiar to those from an optimization background.} sampler for uniformly sampling from $\mc K$.
Each iteration consists of two steps, one that might leave the body and the second accepted only if it is (back) in $\mc K$.

\begin{algorithm}[H]
\hspace*{\algorithmicindent} \textbf{Input:} initial point
$x_0 \sim \pi_{0}$, convex body $\mc K\subset\R^{d}$, iterations $T$, threshold $N$, and $h>0$.

\hspace*{\algorithmicindent} \textbf{Output:} $x_{T+1}$.

\begin{algorithmic}[1] \caption{$\fb$} 
\label{alg:fb-scheme-unif}

\FOR{$i=0,\dotsc,T$}

\STATE Sample $y_{i+1}\sim
\mc N(x_{i},hI_{d})$.\label{line:forward}

\STATE 
Repeat: Sample $x_{i+1}\sim\mc N(y_{i+1},hI_{d})$ until $x_{i+1}\in\mc K$ or \#attempts$_{i}$ $\geq N$ (declare \textbf{Failure}).\label{line:implement-back} 
\ENDFOR
\end{algorithmic}
\end{algorithm}

It might be illuminating for the reader to compare this algorithm to the well-studied $\bw$ (Algorithm~\ref{alg:ball-walk}); each proposed step is a uniform random point in a fixed-radius ball around the current point, and is accepted only if the proposed point is in the body $\mc K$. In contrast, each iteration of $\fb$ is a two-step process, where the first step (Line~\ref{line:forward}) ignores the boundary of the body, and the second step (Line~\ref{line:implement-back}) is accepted only if a proposal $x_{i+1}$ is a feasible point in $\mc K$. We will presently elaborate on the benefits of this variation. 

Each successful iteration of the algorithm, i.e., one that is not declared ``Failure'', can be called a {\em proper} step. We will see that the number of proper steps is directly bounded by isoperimetric constants (such as Poincar\'e and log-Sobolev) of the target distribution. In fact, this holds quite generally without assuming the convexity of $\mc K$. The implementation of an iteration is based on rejection sampling (Line~\ref{line:implement-back}), and our analysis of the efficiency of this step relies crucially on the convexity of $\mc K$. This is reminiscent of the $\sw$ in the literature on convex body sampling (Algorithm~\ref{alg:speedy-walk-uniform}), which is used as a tool to analyze proper steps of the $\bw$. We refer the reader to \S\ref{app:bw-sw} for a brief survey on these and related walks. 

This simple algorithm can be interpreted as a composition of ``flows'' in the space of measures. This view will allow us to use tools from stochastic analysis. 
In particular, we shall demonstrate how to interpret the two steps of one iteration of $\fb$ as alternating \emph{forward} and \emph{backward} heat flows.

We begin by defining an augmented probability measure on $\Rd\times \Rd$ by
\[
\pi(x, y) \propto \exp\bpar{-\frac{1}{2h}\,\abs{x-y}^2}\,\ind_{\mc K}(x)\,.
\]
We denote by $\pi^X, \pi^{X|Y}(\cdot\,|\,y)$ the marginal distribution of its first component (i.e., conditional distribution given the second component), and similarly denote by $\pi^{Y}, \pi^{Y|X}(\cdot\,|\,x)$ for the second component. 
In particular, the marginal in the first component $\pi^X$ is the uniform distribution over $\mc K$. Sampling from such a joint distribution to obtain the marginal on $X$ (say), can be more efficient than directly working only with $\pi^X$. This idea was utilized in Gaussian cooling~\cite{cousins2018gaussian} and later as the restricted Gaussian Oracle (RGO)~\cite{lee2021structured, chen2022improved}. 

Under this notation, Algorithm~\ref{alg:fb-scheme-unif} {without the failure condition} corresponds to a Gibbs sampling scheme from the two marginals of $\pi(x,y)$. 
To be precise, Line~\ref{line:forward} and Line~\ref{line:implement-back} correspond to sampling 
\[
y_{i+1} \sim \pi^{Y|X}(\cdot \,|\, x_i) = \mc N(x_i, h I_d) \qquad \mbox{ and } \qquad x_{i+1} \sim \pi^{X|Y}(\cdot \,|\, y_{i+1}) = \mc N(y_{i+1}, h I_d)|_{\mc K}\,.
\]
It is well known that such a Gibbs sampling procedure will ensure the desired stationarity of $\pi(x,y)$.
We implement the latter step through rejection sampling; if the number of trials in Line~\ref{line:implement-back} hits the threshold $N$, then we halt and declare  \emph{failure} of the algorithm. 
Note that {the accepted point in Line~\ref{line:implement-back}, conditional on success (i.e., number of attempts is at most the cap), is still exactly $\pi^{X|Y=y_{i+1}}$. However, conditioning on ``no failure over all $T$ iterations'' (or equivalently ``re-run-until-success'') re-weights trajectories, so the final output law of the capped INO is not exactly the same as the INO without the cap; see \S\ref{app:bias-from-failure} for details. Nonetheless, it is $\eta$-close to the intended target $\pi^X$ pointwise when the overall failure probability is $\eta$.}

\para{Practical implementation via restart.}
{In practice, one would restart the algorithm if it declares failure before completing the target number of iterations. This yields an always-terminating implementation whose output law equals that of Algorithm~\ref{alg:fb-scheme-unif} conditioned on success. The expected number of restarts is at most $(1-\eta)^{-1}$, so the expected total query complexity increases by a multiplicative factor of only $(1-\eta)^{-1} \le 2$ for $\eta \le 1/2$.}

\para{Stochastic perspective: heat flows.}
Our algorithm can be viewed through the lens of stochastic analysis of proximal sampling \cite{chen2022improved}.
This view provides an interpolation in continuous-time, which is simple and powerful. 
To make this concrete, we follow the exposition from \cite[\S 8.3]{chewi2023log}. We denote the successive laws of $x_i$ and $y_i$ by $\mu^{X}_i$ and $\mu^Y_i$, respectively. 
Recall that the first step of sampling from $\pi^{Y|X}(\cdot | x_i)$ (Line~\ref{line:forward}) yields $\mu^Y_{i+1} = \int \pi^{Y|X = x}\,\D \mu^X_i(x)$. This is the result of evolving a probability measure under \emph{(forward) heat flow} of $\mu_i^X$ for some time $h$, given by the following stochastic differential equation: for $Z_0 \sim \mu_i^X$,
\begin{equation}
    \D Z_t = \D B_t
    \tag{$\msf{FH}$}\label{eq:forward-heat}
\end{equation}
where $B_t$ is the standard Brownian process.
We write $\law(Z_t) = \mu_i^X P_t$. In particular, $Z_h = Z_0 + \zeta \sim \mu_i^X * \mc N(0,hI_d) = \mu_{i+1}^Y$ for $\zeta \sim \mc N(0, hI_d)$.
When $\mu_i^X = \pi^X$, the first step of Algorithm~\ref{alg:fb-scheme-unif} gives
\begin{equation}
    \pi^Y(y) = [\pi^X * \mc N(0, hI_d)](y) = \frac{1}{\vol(\mc K)\,(2\pi h)^{d/2}} \int_{\mc K}\exp \bpar{-\frac{1}{2h}\abs{y-x}^2} \,\D x\,.\label{eq:unif-conv}
\end{equation}

The second step of sampling from $\pi^{X|Y}(\cdot\, |\,y_{i+1})$ can be represented by $\mu_{i+1}^X = \int \pi^{X|Y=y}\,\D\mu_{i+1}^Y(y)$ (Line~\ref{line:implement-back}).
Although we do not require it in the analysis, it is interesting to note that the backward step can also be viewed as a continuous-time process.
Consider \eqref{eq:forward-heat} with $Z_0 \sim \pi^X$. Then, $Z_h \sim \pi^Y$, so the joint distribution of $(Z_0, Z_h)$ is simply $\pi$. 
This implies that $(Z_0 | Z_h = y) \sim \pi^{X|Y=y}$.
Imagine there is an SDE \emph{reversing} the forward heat flow in a sense that if we initialize deterministically at $Z_h \sim \delta_y$ at time $0$, then the law of the SDE at time $h$ would be $\law(Z_0 | Z_h=y) = \pi^{X|Y=y}$.
Then, this SDE would serve as a continuous-time interpolation of the second step. 

\begin{table}
\begin{centering}
\begin{tabular}{c|c|c}
\toprule 
 &$\qquad$ Forward flow$\qquad$ & Backward flow\tabularnewline
\midrule
\midrule 
SDE & $\D Z_{t}=\D B_{t}$ & $\D Z_{t}^{\leftarrow}=\nabla\log(\pi^{X}P_{h-t})(Z_{t}^{\leftarrow})\,\D t+\D B_{t}$\tabularnewline
\midrule 
Fokker--Planck & $\de_{t}\mu_{t}=\half\Delta\mu_{t}$ & $\de_{t}\mu_{t}^{\leftarrow}=-\Div\bpar{\mu_{t}^{\leftarrow}\nabla\log(\pi^{X}P_{h-t})}+\half\Delta\mu_{t}^{\leftarrow}$\tabularnewline
\bottomrule
\end{tabular}
\par\end{centering}
\caption{The Fokker--Planck equations for the forward and backward heat flow describe how the laws of $Z_t$ and $Z_t^\leftarrow$ in \eqref{eq:forward-heat} and \eqref{eq:backward-heat} evolve over time. See \S\ref{app:contraction-sketch} for details.}
\end{table}

Such a {\em time reversal} SDE is indeed possible! The following SDE on $(Z_t^\leftarrow)_{t\in [0,h]}$ initialized at $Z_0^\leftarrow \sim \pi^Y = \pi^X P_h$ ensures 
$Z_{h-t} \sim \law(Z_t^\leftarrow) = \pi^XP_{h-t}$:
\begin{equation}
    \D Z_t^\leftarrow = \nabla \log (\pi^X P_{h-t}) (Z_t^\leftarrow)\,\D t + \D B_t\quad \text{for } t\in[0,h]\,.\tag{$\msf{BH}$}\label{eq:backward-heat}
\end{equation}
Although this is designed to reverse \eqref{eq:forward-heat} \textbf{initialized by $Z_0 \sim \pi^X$} (so $Z_h = Z_0^\leftarrow \sim \pi^Y)$, 
its construction also ensures that if $Z_0^\leftarrow \sim \delta_{y}$, a point mass, then $Z_h^\leftarrow \sim \law(Z_0 | Z_h = y) = \pi^{X | Y = y}$.
Thus, if we initialize \eqref{eq:backward-heat} with $Z_0^\leftarrow \sim \mu_{i+1}^Y$, then the law of $Z_h^\leftarrow$ corresponds to $\int \pi^{X|Y=y}\,\D \mu_{i+1}^Y(y) = \mu_{i+1}^X$. 
We mention this reverse process due to its elegant symmetry but it will not be needed in our mixing analysis.

\begin{remark}\label{rmk:proximal}
We note that $\fb$ is exactly the {\em proximal} sampling scheme \cite{lee2021structured, chen2022improved, fan2023improved} for uniform distributions. The proximal sampler with a target density proportional to $\exp(-V)$ considers an augmented distribution $\pi(x, y) \propto \exp(-V(x)-\frac{1}{2h}\,\abs{x-y}^2)$ and then repeats the following two steps: (1) $y_{i+1} \sim \pi^{Y|X=x_i} = \mc N(x_i, hI_d)$ and then  (2) $x_{i+1} \sim \pi^{X | Y=y_{i+1}}$.
Na\"ively, the proximal sampler is implemented by performing rejection sampling, with the Gaussian centered at the minimizer of $\log \pi^{\cdot | Y=y_{i+1}}$ as the proposal. Realizing this would require a projection oracle (to $\mc K$), which is only known to be implementable with $O(d^2)$ membership queries. $\fb$ completely avoids the need for a projection oracle.
\end{remark}

\subsection{Results}\label{sec:results}
Our computational model is the classical general model for convex bodies~\cite{grotschel2012geometric}. We assume $\vol \mc K >0$ in this paper. Below, $B_r(x)$ denotes the $d$-dimensional ball of radius $r$ centered at $x$.

\begin{definition}[Convex body oracle]\label{def:membership}
    A \emph{well-defined membership oracle} for a convex body $\mc K \subset \Rd$ is given by a point $x_0 \in \mc K$, a number $D > 0$, with the guarantee that $B_1(x_0) \subseteq \mc K \subseteq B_D(x_0)$, and an oracle that correctly answers \emph{YES} or \emph{NO} to any query of the form ``$x \in \mc K$?'' 
\end{definition}

\begin{definition}[Warmness]\label{as:warmness}
    A distribution $\mu$ is said to be \emph{$M$-warm} with respect to another distribution $\pi$ if for every $x$ in the support of $\pi$, we have  $\D \mu(x) \le M\, \D \pi(x)$.
\end{definition}

We now summarize our main result, which is further elaborated in \S\ref{sec:example-main-result}.
Below, $\pi^{\mc K}$ is the uniform distribution over $\mc K$, and $\eu R_q$ is the R\'enyi-divergence of order $q$ (see Definition~\ref{def:p-dist}).
\begin{theorem}[Succinct version of Theorem~\ref{thm:main-result}]
For any given $\eta, \varepsilon \in (0,1/2)$, $q\geq 2$, and any convex body $\mc K$ given by a well-defined membership oracle, there exist choices of parameters $h, N$ such that $\fb$,  starting from an $M$-warm distribution, with probability at least $1-\eta$, returns $X\sim \mu$ such that $\eu R_q(\mu \mmid \pi^{\mc K}) \leq \varepsilon + 4\eta$. The number of proper steps is $\Otilde(q d^2 \Lambda \log^2\nicefrac{M}{\eta \veps})$, and the expected total number of membership queries is $\Otilde(qM d^2 \Lambda \log^6\nicefrac{1}{\eta \veps})$, where $\Lambda$ is the largest eigenvalue of the covariance of $\pi^{\mc K}$. 
\end{theorem}

Even though our guarantee is in the much stronger ``metric'' of $\eu R_q$ compared to the $\tv$ guarantees of the $\bw$, we do not incur any additional asymptotic complexity.
To obtain this result, we choose the following values for the parameters: $h^{-1} = \widetilde{\Theta}(d^2 \log \frac{qM\Lambda}{\eta} \log \log \frac{1}{\varepsilon})$, $N = \widetilde{\Theta}(\eta^{-1}qM d^2 \Lambda \log^5(1/\varepsilon))$. See Lemma~\ref{lem:per-iteration-guarantees} for more details. The complexity of the restart-on-failure algorithm immediately follows from this result as a simple corollary.
{
\begin{corollary}[Restart-on-failure]\label{cor:comp-restart}
For $\veps\in (0, \frac{1}{10})$ and $q\geq 2$, consider the restart variant that repeats Algorithm~\ref{alg:fb-scheme-unif} from scratch until success with the same parameter choices. Then, the output sample $X \sim \mu'$ satisfies $\eu{R}_q(\mu'\,\|\,\pi^{\mc{K}})\leq \veps$, and the expected total number of membership queries is $\Otilde(qM d^2 \Lambda \log^6\nicefrac{1}{\veps})$.
\end{corollary}
}

\begin{remark}[(Warm-start generation)]
    While the assumption of warmness for the initialization may seem strong at the outset, for well-rounded convex bodies ($\E_{\pi}[\abs{\cdot}^2] \leq C^2 d$ for some constant $C$), it is possible to generate an $\O(1)$ warm-start with complexity $\Otilde(d^3)$. We refer readers to \cite{cousins2018gaussian, kook2024r, kook2025faster} for details.
\end{remark}

We note that for $X\sim \pi^{\mc K}$,
\[
\norm{\cov \pi^{\mc K}}_{\msf{op}} \leq \tr(\cov \pi^{\mc K}) = \E[\abs{X-\E X}^2] \leq D^2\,.
\]

The above guarantee in the R\'enyi divergence immediately provides $\mc W_2, \msf{TV}, \KL$, and $\chi^2$ guarantees as special cases.
For two distributions $\mu$ and $\pi$, we have
\begin{enumerate}
\item $\KL(\mu \mmid \pi) = \lim_{q\downarrow 1} \eu R_q(\mu \mmid \pi) \leq \eu R_q(\mu \mmid \pi) \leq \eu R_{q'}(\mu \mmid \pi) \leq \eu R_{\infty}(\mu \mmid \pi) = \log \sup\frac{\D\mu}{\D\pi}$ for  $1<q\leq q'$.
\item $2\,\norm{\mu -\pi}_{\msf{TV}}^2\leq \KL(\mu\mmid \pi)\leq \log(\chi^2(\mu \mmid \pi)+1) = \eu R_2(\mu \mmid \pi)$.
\item $\mc W_2^2(\mu, \pi) \leq 2C_{\msf{LSI}}(\pi)\, \KL(\mu \mmid \pi)$ (Talagrand's $\msf{T}_2$-inequality)  and $C_{\msf{LSI}}(\pi^{\mc K}) \lesssim D^2$.
\item $\mc W_2^2(\mu, \pi) \leq 2C_{\msf{PI}}(\pi)\, \chi^2(\mu \mmid \pi)$ \cite{liu2020poincare} and $C_{\msf{PI}}(\pi^{\mc K})\lesssim \norm{\cov\pi^{\mc K}}_{\msf{op}} \log d$.
\end{enumerate}

The query complexity is better if the convex body is (near-)\emph{isotropic}, i.e., the uniform distribution over the body has (near-)identity covariance. This relies on recent estimates of the worst-case Poincar\'e constant for isotropic logconcave distributions~\cite{kannan1995isoperimetric,klartag2023logarithmic}. The condition that the convex body is isotropic can be achieved in practice through a \emph{rounding} procedure~\cite{jia2021reducing,JLLV24reducing, kook2024covariance}. See \S\ref{sec:related} for more details.

\begin{corollary}\label{cor:isotropic-mixing}
Assume that $\pi^{\mc K}$ is near-isotropic, i.e., the operator norm of its covariance is $\O(1)$. Under the same setting as above, $\fb$ succeeds with probability $1-\eta$, returning $X\sim \mu$ such that $\eu R_q(\mu \mmid \pi^{\mc K}) \leq \veps + 4\eta$.
The number of proper steps is $\Otilde(qd^2\log^2\frac{M}{\eta \veps})$, and the expected total number of membership queries is $\Otilde(qMd^2 \log^6\tfrac{1}{\eta \veps})$.
\end{corollary}

{Building on the contraction results in \cite{chen2022improved,KO25strong}, we show that} the bound on the number of proper steps holds for general \emph{non-convex} bodies and from \emph{any feasible start} in $\mc K$.
We remark that this bound for non-convex uniform sampling is not known for the $\bw$ or the $\sw$.

\begin{corollary}\label{thm:any-start}
For any given $\varepsilon\in(0,1)$ and set $\mc K \subset B_D(0)$, $\fb$ with variance $h$ and any feasible start $x_0\in \mc K$ achieves $\eu R_q(\mu^X_m \mmid \pi^{X}) \leq \veps$ after $m = \Otilde(qh^{-1}C_{\msf{LSI}}(\pi^X)\log\frac{d+D^2/h}{\veps})$ iterations.
\end{corollary}

\begin{corollary}\label{cor:any-start-convex}
For any given $\varepsilon\in(0,1)$ and convex body $\mc K\subset B_D(0)$, $\fb$ with variance $h$ and a feasible start $x_0 \in \mc K$ achieves $\eu R_q(\mu^X_m \mmid \pi^X)\leq \veps$ after $m=\Otilde(qh^{-1}D^2\log\frac{1}{\veps})$ iterations. If $\pi^{X}$ is isotropic, then $\fb$ only needs $\Otilde(qh^{-1}D\log\frac{d+d^2/h}{\veps})$ iterations.
\end{corollary}

{We note that, with the step size $h\asymp d^{-2}$, Corollary~\ref{cor:any-start-convex} immediately recovers mixing guarantees matching the ones for the $\sw$ obtained via blocking conductance and the log-Cheeger inequality \cite{kannan2006blocking} (see Theorem~\ref{thm:sw-results}).}

\para{Outline of analysis.}
We summarize our proof strategy below, which requires us to demonstrate two facts: (i) The current distribution should converge to the uniform distribution, (ii) within each iteration of the algorithm, the failure probability and the expected number of rejections should be small enough.

\begin{itemize}
    \item 
    We need to demonstrate that the corresponding Markov chain is rapidly mixing. 
    Here, we use the heat flow perspective to derive mixing rates under any suitable divergence measure (such as $\KL$, $\chi^2$, or $\eu R_q$). This extends known results for the unconstrained setting~\cite{chen2022improved}. To summarize the proof, by considering instead the solutions after small time $t$, we invoke known contraction results from~\cite{chen2022improved} and then use a continuity argument to conclude the proof.

    \item We show that the number of rejections in Line~\ref{line:implement-back} over the entire execution of the algorithm is bounded with high probability. To do this, we apply a detailed argument involving local conductance and the convexity of $\mc K$, which relies on techniques from~\cite{belkin2006heat}. For this step, we show that with the appropriate choice of variance $h = \widetilde\Theta(d^{-2})$ and threshold $N = \widetilde\Theta(T\eta^{-1})$, the entire algorithm succeeds with probability $1-\eta$. The expected number of rejections is polylogarithmic.
\end{itemize}

While each individual component resembles pre-existing work in the literature, in their synthesis we will demonstrate how to interleave past relevant developments in theoretical computer science, optimal transport, and functional analysis. 
This yields elegant and surprisingly simple proofs, as well as stronger results.

\subsection{Discussion}\label{sec:remark-result} 
\textbf{No need to be lazy.\,}
Previous uniform samplers like the $\bw$ are made  \emph{lazy} (i.e., with probability $1/2$, it does nothing), to ensure convergence to the target stationary distribution. However, our algorithm does not need this, as our sampler is shown to directly contract towards the target.

\para{Unified framework.} 
We remark that Theorem~\ref{thm:conv-lsi-pi} places the previously known mixing guarantees for the $\bw, \sw$ in a unified framework.
Existing tight guarantees for the $\sw$ are in TV distance and based on the log-Sobolev constant, assuming an oracle for implementing each step~\cite{lee2017eldan}. 
The known convergence guarantees of the $\bw$ (see \S\ref{app:bw-sw} for details), namely the mixing time of $\Otilde(Md^2D^2\log\frac{1}{\veps})$ for TV distance, are for the composite algorithm [$\sw + $rejection sampling]. Here, the $\sw$ records only the accepted steps of the $\bw$, so its stationary distribution differs slightly from the uniform distribution (and can be corrected with a post-processing step).
On the other hand, $\fb$ actually converges to $\pi^{\mc K}$ without any adjustments and achieves stronger R\'enyi divergence bounds in the same asymptotic complexity. 
Our analysis shows that the mixing guarantee is determined by isoperimetric constants of the target  (Poincar\'e or log-Sobolev). 

\para{Effective step size.\,} 
The $\bw$'s largest possible step size is of order $1/\sqrt d$ (see  \S\ref{app:bw-sw}) to keep the rejection probability bounded by a constant.
This bound could also be viewed as an ``effective'' step size of $\fb$, since the $\ell_2$-norm of the Gaussian $\mc N(0,hI)$ is concentrated around $\sqrt{hd}$ and we will set the variance $h$ of $\fb$ to $\Otilde(d^{-2})$, so we have $\sqrt{hd} \approx 1/\sqrt{d}$.

\para{What has really changed?} 
$\fb$ has clear similarities to both $\bw$ and $\sw$. 
What then are the changes that allow us to use  continuous-time interpolation? 
One step of the $\bw$ is [random step ($y\in B_{\delta}(x)$) $+$ Metropolis-filter (accept if $y\in \mc K$)].
This filtering is an abrupt discrete step, and it is unclear how to control contraction. It could be replaced by a step of the $\sw$ ($x\sim \text{Unif}(B_{\delta}(y)\cap \mc K)$). Then, each iteration of $\fb$ can be viewed as a Gaussian version of a $\msf{Ball\ walk's\ proposal} + \sw$ algorithm.

How can we compare $\fb$ with the $\sw?$ Iterating speedy steps leads to a biased distribution. 
One step of (a Gaussian version of) the $\sw$ can be understood as a step of backward heat flow. Therefore, if one can control the isoperimetric constants of the biased distribution along the trajectory of the backward flow, then contraction of the $\sw$ toward the biased distribution will follow from the simultaneous backward analysis.

\para{Subsequent work.}
The ideas and methods of this paper have led to subsequent progress. First, \cite{kook2024r} shows that the output guarantee can be strengthened to $\eu{R}_\infty$ under LSI, using $\Otilde(d^2R^2\polylog\frac{1}{\veps\eta})$ queries.  
Second, the diffusion-based approach is extended to general logconcave sampling in \cite{KV25sampling}, with the query complexity of $\Otilde(d^2\,(R^2\vee 1) \polylog\frac{1}{\veps\eta})$ for a $\eu{R}_\infty$ guarantee, beyond the uniform distribution. Lastly, the $\eu R_\infty$-warmness requirement has been relaxed to significantly weaker $\eu R_{\Otilde(1)}$-warmness in \cite{kook2025faster}, improving the query complexity of warm-start generation from $\Otilde(d^2R^2)$ to $\Otilde(d^2R^{3/2}\norm{\cov \pi}^{1/4})$.

\subsection{Related work}\label{sec:related}
Sampling from constrained logconcave distributions is a fundamental task arising in many fields. Uniform sampling with convex constraints is its simplest manifestation, which was first studied as a core subroutine for a randomized volume-computation algorithm \cite{dyer1991random}. Since then, this fundamental problem has been studied for over three decades \cite{lovasz1990mixing,lovasz1993random,kannan1997random,lovasz2006hit, bubeck2018sampling,pmlr-v65-brosse17a}. We review these algorithms, grouping them under three categories --- geometric random walks, structured samplers, and diffusion-type samplers. Below, $\mc K$ is convex.

\para{Geometric random walks.}
We discuss two geometric random walks -- $\bw$ \cite{lovasz1993random,kannan1997random} and $\har$ \cite{smith1984efficient,lovasz1999hit}. The $\bw$ is a simple metropolized random walk; it draws $y$ uniformly at random from a ball of radius $\delta$ centered at a current point $x$, and moves to $y$ if $y\in \mc K$ and stays at $x$ otherwise. In the literature, the $\bw$ actually refers to a composite algorithm consisting of [$\sw +$ rejection sampling], where the $\sw$ records only the accepted steps of the $\bw$ (see \S\ref{app:bw-sw} for details).
The step size $\delta$ should be set to $\O(d^{-1/2})$ to avoid stepping outside of $\mc K$. 
\cite{kannan1997random} showed that the $\bw$ needs $\Otilde(Md^2D^2\log\frac{1}{\veps})$ membership queries to be $\veps$-close to $\pi^{\mc K}$ in $\tv$, where $D$ is the diameter of $\mc K$, and the warmness parameter $M$ measures the closeness of the initial distribution to the target uniform distribution $\pi^{\mc K}$.

$\har$ is another zeroth-order algorithm that needs \emph{no step size}; it picks a uniform random line $\ell$ passing a current point, and move to a uniform random point on $\ell \cap \mc K$. 
\cite{lovasz2006hit} shows that, if we define the second moment as $R^2 := \E_{\pi^{\mc K}}[\abs{X-\E X}^2]$, then $\har$ requires $\O(d^2R^2\log\frac{M}{\veps})$ iterations. Notably, this algorithm has a poly-logarithmic dependence on $M$ as opposed to the $\bw$. Chen and Eldan \cite{CE25hitandrun} further showed that $\har$ mixes in $\Otilde(M^{11}d^2/\veps^{11})$ times for isotropic convex bodies. 

Both algorithm are affected by skewed shape of $\mc K$ (i.e., large $D$ or $R$), so these samplers are combined with pre-processing step called \emph{rounding}. This procedure finds a linear transformation that makes the geometry of $\mc K$ less skewed and so more amenable to sampling. In literature, there exists a randomized algorithm \cite{jia2021reducing} that rounds $\mc K$ and generates a good warm start (i.e., $M=\O(1)$), with the $\bw$ used as a core subroutine. We refer readers to \cite{kook2024covariance} for a streamlined proof, with  $\fb$ used as a main sampler.
This algorithm takes up $\Otilde(d^{3.5})$ queries in total, and from a well-rounded position with a good warm start, the $\bw$ only needs $\Otilde(d^2\log\frac{1}{\veps})$ queries to sample from $\pi^{\mc K}$.

\para{Structured samplers.}
The aforementioned samplers based on geometric random walks require \emph{only} access to the membership oracle of the convex body \emph{without} any additional structural assumptions. The alternate paradigm of \emph{geometry-aware sampling} attempts to exploit the \emph{structure} of convex constraints, with the aim of expediting the convergence of the resultant sampling schemes. One common assumption is to make available a \emph{self-concordant barrier function} $\phi$ which has regularity on its high-order derivatives and blows up when approaching the boundary $\de \mc K$. The Hessian of $\phi$ encodes the local geometry of the constraint, and the samplers often work directly with $\hess \phi$.

The first canonical example of such a zeroth-order sampler is $\dw$ used when $\mc K$ is given by $m$ linear constraints~\cite{kannan2012random}; it draws a uniform sample from an ellipsoid (characterized by $\hess \phi$) of fixed radius around a current point, and is often combined with a Metropolis adjustment. \cite{kannan2012random} shows that $\dw$ mixes in $\O(md\log\frac{M}{\veps})$ steps, although each iteration is slightly more expensive than one membership query. This algorithm requires no rounding, but still needs a good warm-start, which can be achieved by an annealing-type algorithm using $\Otilde(md)$ iterations of $\dw$ \cite{kook2024gaussian}.

$\rhmc$ is a structured sampler that exploits the first-order information of the potential (i.e., $\nabla \log (1/\pi)$) \cite{girolami2011riemann}; its proposal is given as the solution to the Hamilton's ODE equation, followed by the Metropolis-filter. In the linear-constraint setting above, it requires $\O(md^{2/3}\log\frac{M}{\veps})$ many iterations to achieve $\veps$-close distance to $\pi^{\mc K}$ \cite{lee2018convergence}. This sampler is further analyzed for practical ODE solvers \cite{kook2022condition} and for more sophisticated self-concordant barriers \cite{gatmiry2023sampling}.

Similarly, $\ml$ \cite{zhang2020wasserstein, jiang2021mirror, ahn2021efficient, li2022mirror} is a class of algorithms which converts the constrained problem into an unconstrained one obtained by considering the pushforward of the constrained space by $\nabla \phi$. The algorithm can also be metropolized~\cite{srinivasan2023fast}. The best known rate for this algorithm is $\widetilde{O}(d\log \frac{1}{\varepsilon})$ under some strong assumptions on $\phi$.

\para{Diffusion-based samplers.}
Samplers based on discretizations of It\^o diffusions, stochastic processes which rapidly mix to $\pi$ in continuous time, have long been used for sampling without constraints~\cite{besag1995bayesian, dalalyan2012sparse, dalalyan2017further, chewi2023log}. While the underlying stochastic processes generalize easily to constrained settings, the discretization analysis relies crucially on the smoothness of the target distribution. This is clearly impossible to achieve in the constrained setting, so some techniques are required to circumvent this difficulty. These algorithms, however, generalize easily to the more general problem of sampling from distributions of the form $\tilde \pi^X \propto e^{-f} \ind_{\mc K}$, by incorporating first-order information from $f$.

The first approach for adapting diffusion-based samplers~\cite{bubeck2018sampling, lehec2023langevin} iterates a two-step procedure. First, a random step is taken, with $x_{k+1/2} \sim \mc N(x_{k}, 2hI_d)$ for some appropriately chosen step $h$,\footnote{A gradient step can be added in the more general case, for sampling from $\tilde \pi^X$.} and then project it to $\mc K$, i.e., $x_{k+1} = \msf{proj}_{\mc K}(x_{k+1/2})$. The complexity is given in terms of queries to a \emph{projection oracle}, each call to which can be implemented with a polynomial number of membership oracle queries; a total of $\Otilde(\nicefrac{d^2 D^3}{\varepsilon^4})$ queries are needed to be $\varepsilon$-close in $\mc W_2$ to $\pi^X$. Another approach, which uses an algorithmically designed ``soft'' penalty instead of a projection, was proposed in~\cite{gurbuzbalaban2022penalized}, and achieves a rate estimate of $\Otilde(\nicefrac{d}{\varepsilon^{10}})$. 

A second approach, suggested by~\cite{pmlr-v65-brosse17a}, considers a different proximal scheme, which performs a ``soft projection'' onto $\mc K$, by taking steps like $\mc N((1-h\lambda^{-1})x_{k} + h\, \msf{proj}_{\mc K}(x_k), 2hI_d)$. It is called Moreau--Yosida regularized Langevin, named after an analogous regularization scheme for constrained optimization. This scheme also relies on access to a projection oracle for $\mc K$, and quantifies their query complexity accordingly. Their final rate estimate is $\Otilde(\nicefrac{d^5}{\varepsilon^6})$ to be $\varepsilon$-close in $\tv$ distance to $\pi^X$.

Observing the prior work integrating diffusion-based sampling with convex constraints, the dependence on the key parameters $d, \varepsilon$, while polynomial, are many orders worse than the rates for zeroth-order samplers such as $\bw$ and $\har$. In contrast, our analysis not only recovers but in some sense surpasses the known rates for $\bw$ and $\har$, while harmonizing well with the continuous-time perspective of diffusions. 

\para{Proximal schemes for sampling.}
The Gibbs sampling scheme used in this paper was inspired by the restricted Gaussian oracle introduced in~\cite{lee2021structured} (in turn inspired by Gaussian cooling~\cite{cousins2018gaussian}), which alternately iterates between a pure Gaussian step, and a ``proximal'' step (which we elaborate in our exposition). This scheme was given novel interpretations by~\cite{chen2022improved}, which showed that it interpolates the forward and backward heat flows, in the sense defined by~\cite{klartag2021spectral}. The backward heat flow itself is intimately related to stochastic localization schemes, invented and popularized in~\cite{eldan2013thin, chen2021almost}.

This formulation proved surprisingly powerful, allowing many existing rates in unconstrained sampling to be recovered from a relatively simple analysis. This was further extended by~\cite{fan2023improved} to achieve the current state-of-the-art rate in unconstrained sampling. Finally, \cite{gopi2023algorithmic} suggests that this could be applied to tackle some constrained problems.
However, the assumptions in this final mentioned work are not compatible with the uniform sampling problem on general convex bodies.

\section{Preliminaries}
Unless otherwise specified, we will use $\abs{\cdot}$ for the $2$-norm on $\R^d$. We write $a = O(b)$, $a \lesssim b$ to mean that $a \leq cb$ for some universal constant $c>0$. Similarly, we write $a \gtrsim b, a = \Omega(b)$ for $a \geq c b$, while $a = \Theta(b)$ means  $a \lesssim b, b \lesssim a$ simultaneously. We will also use $a = \Otilde(b)$ to denote $a = O(b \polylog b)$. Lastly, we will use measure and density interchangeably when there is no confusion.

To quantify the convergence rate, we recall common notions of divergence between distributions.
\begin{definition}[Distance and divergence] \label{def:p-dist}
    For two measures $\mu, \nu$ on $\R^d$, the \emph{total variation} distance between them is defined by
    \[
        \norm{\mu-\nu}_{\msf{TV}} := \sup_{B \in \mc F}\, \abs{\mu(B) - \nu(B)}\,,
    \]
    where $\mc F$ is the collection of all measurable subsets of $\R^d\,$. 
    The $2$-\emph{Wasserstein distance} is given by
    \[
        \mc W_2^2(\mu, \nu) := \inf_{\gamma \in \Gamma(\mu, \nu)} \E_{(X, Y) \sim \gamma} [\abs{X-Y}^2]\,,
    \]
    where $\Gamma$ is the set of all couplings between $\mu, \nu$. 
    Next, we define the \emph{$f$-divergence} of $\mu$ towards $\nu$ with $\mu \ll \nu$ (i.e., $\mu$ is absolutely continuous with respect to $\nu$) as, for some convex function $f:\R_+ \to \R$ with $f(1) = 0$ and  $f'(\infty) = \infty$,
    \[
        D_f(\mu \mmid \nu) := \int f\bpar{\frac{\D \mu}{\D \nu}} \, \D \nu\,.
    \]
    The \emph{$\KL$-divergence} arises when taking $f(x) = x \log x$, the \emph{$\chi^q$-divergence} when taking $f(x) = x^q - 1$, and the \emph{$q$-R\'enyi divergence} is given by
    \[
        \eu R_q(\mu \mmid \nu) := \frac{1}{q-1} \log \bpar{\chi^q(\mu \mmid \nu) + 1}\,.
    \]
\end{definition}

We recall two important functional inequalities of a distribution.

\begin{definition}
We say that a probability measure $\nu$ on $\R^d$ satisfies a \emph{Poincar\'e inequality} (PI) with parameter $C_{\msf{PI}}(\nu)$ if for all smooth functions $f: \R^d \to \R$,
\[
\var_{\nu}f \leq C_{\msf{PI}}(\nu)\, \E_{\nu}[\abs{\nabla f}^2]\,,\label{eq:pi}\tag{$\msf{PI}$}
\]
where $\var_\nu f \deq \E_\nu [\abs{f - \E_\nu f}^2]$.
\end{definition}

The Poincar\'e inequality is implied by the log-Sobolev inequality.

\begin{definition}
We say that a probability measure $\nu$ on $\R^d$ satisfies a \emph{log-Sobolev inequality} (LSI) with parameter $C_{\msf{LSI}}(\nu)$ if for all smooth functions $f:\Rd \to \R$,
\[
\Ent_\nu(f^2) \leq 2C_{\msf{LSI}}(\nu)\, \E_\nu[\abs{\nabla f}^2]\,, \label{eq:lsi}\tag{$\msf{LSI}$-$\msf{I}$}
\]    
where $\Ent_\nu(f^2) := \E_\nu[f^2 \log f^2] - \E_\nu[f^2] \log (\E_\nu[f^2])$.
Equivalently, for any probability measure $\mu$ over $\Rd$ with $\mu\ll\nu$,  
\[
\KL(\mu \mmid \nu) \leq \frac{C_{\msf{LSI}}(\nu)}2\, \FI(\mu \mmid \nu)\,,\label{eq:lsi2}\tag{$\msf{LSI}$-$\msf{II}$}
\]
where $\FI(\mu \mmid \nu):= \E_{\mu}[\abs{\nabla \log\frac{\D\mu}{\D\nu}}^2]$ is the \emph{Fisher information} of $\mu$ with respect to $\nu$.
\end{definition}

We state two important lemmas which are needed for our proofs. The first is the data-processing inequality (DPI) for R\'enyi divergence and $f$-divergence, given below.

\begin{lemma}[Data-processing inequality]\label{lem:DPI}
    For measures $\mu, \nu$, Markov kernel $P$, $f$-divergence $D_f$, and $q \geq 1$, it holds that
    \[
        D_f(\mu P \mmid \nu P) \leq D_f (\mu  \mmid \nu )\,,\quad\text{and}\quad
        \eu R_q(\mu P \mmid \nu P) \leq \eu R_q (\mu \mmid \nu)\,.
    \]
\end{lemma}

Functional inequalities allow us to show exponential contraction of various divergences, through the following helpful inequality.
\begin{lemma}[Gr\"onwall]\label{lem:gronwall}
    Suppose that $u, g: [0, T] \to \R$ are two continuous functions, with $u$ being differentiable on $[0,T]$ and satisfying
    \[
        u'(t) \leq g(t)\, u(t) \qquad \text{for all } t \in [0,T]\,.
    \]
    Then,
    \[
        u(t) \leq \exp\Bpar{\int_0^t g(s) \, \D s}\, u(0) \qquad  \text{for all } t \in [0,T]\,.
    \]
\end{lemma}

\section{Analysis}
We begin this section by proving the stationarity of the target $\pi^X$.

\begin{lemma}
$\pi^X$ is stationary under iterations of $\fb$.
\end{lemma}

\begin{proof}
    Note that the transition kernel of the forward and backward step is $\pi^{Y|X=\cdot}$ and $\pi^{X|Y=\cdot}$, respectively. Hence, the forward step brings $\pi^X$ to $\pi^Y$ as seen in
    \[
     \int \pi^{Y|X}(\cdot\,|\,x)\, \pi^X(x)\,\D x
    =\int \pi^{X,Y}(x,\cdot) \,\D x
    = \pi^Y\,.
    \]
    Also, the backward step brings $\pi^Y$ to $\pi^X$:
    \[
     \int \pi^{X|Y}(\cdot\,|\,y)\, \pi^Y(y)\,\D y
    =\int \pi^{X,Y}(\cdot, y) \,\D y
    = \pi^X\,.
    \]
    Therefore, $\pi^X$ is stationary under each iteration of $\fb$.
\end{proof}

Our analysis for $\fb$ consists of two parts: (1) demonstrating its mixing, i.e., how many \emph{outer} iterations are needed to be sufficiently close to the uniform distribution, and (2) quantifying the failure probability and wasted steps in Line~\ref{line:implement-back}.

For (1), we collect in \S\ref{sec:func-ineq} some important implications of functional inequalities, e.g., the  Poincar\'e and log-Sobolev inequalities, for the uniform distribution. Then in \S\ref{sec:contraction-analysis}, we exploit the flow perspective of the algorithm to obtain the mixing guarantees. To this end, we revisit the proofs for the contraction results of forward heat flows in \cite{chen2022improved,KO25strong}.

\begin{theorem}\label{thm:cont-uni-dist}
\phantomsection
    Let $\mu^X_k$ be the law of the $k$-th output of $\fb$ with initial distribution $\mu_0^X$ and step size $h>0$.
    Let $C_{\msf{LSI}}$ be the \eqref{eq:lsi} constant of the uniform distribution $\pi^X$ over $\mc K$. 
    Then, for any $q \geq 1$,
    \[
    \eu R_q(\mu_k^X\mmid \pi^X) \leq \frac{\eu R_q(\mu_0^X \mmid \pi^X)}{(1+h/C_{\msf{LSI}})^{k/q}}\,.
    \]

    For $C_{\msf{PI}}$ the \eqref{eq:pi} constant of $\pi^X$,
    \[
    \chi^2(\mu^X_k \mmid \pi^X) \leq \frac{\chi^2(\mu^X_0 \mmid \pi^X)}{(1+h/C_{\msf{PI}})^{k}}\,.
    \]
    Furthermore, for any $q \geq 2$,
    \[
    \eu R_{q}(\mu_{k}^{X}\mmid\pi^{X})\leq\begin{cases}
    \eu R_{q}(\mu_{0}^{X}\mmid\pi^{X})-\frac{k\log(1+h/C_{\msf{PI}})}{q} & \text{if }k\leq\frac{q}{2\log(1+h/C_{\msf{PI}})}\,\bpar{\eu R_{q}(\mu_{0}^{X}\mmid\pi^{X})-1}\,,\\
    (1+h/C_{\msf{PI}})^{-(k-k_{0})/q} & \text{if }k\geq k_{0}:=\lceil\frac{q}{2\log(1+h/C_{\msf{PI}})}\,\bpar{\eu R_{q}(\mu_{0}^{X}\mmid\pi^{X})-1}\rceil\,.
    \end{cases}
    \]
\end{theorem}

The result reduces the problem of obtaining a mixing guarantee to that of demonstrating a functional inequality on the target distribution. For this, it is not strictly necessary that $\mc K$ be convex.

As for (2), convexity of $\mc K$ is crucial this time unlike (1). We show in \S\ref{sec:fail-prob} that the failure probability remains under control by taking a suitable variance $h$ and threshold $N$, and that the expected number of trials per iteration is of order $\log N$, not $N$. 

\begin{lemma}[Per-iteration guarantees]\label{lem:per-iteration-guarantees}
Let $\mc K$ be any convex body in $\Rd$ presented by a well-defined membership oracle, $\pi^X$ the uniform distribution over $\mc K$, and $\mu$ an $M$-warm initial distribution with respect to $\pi^X$. For any given $m\in \mathbb{N}$ and $\eta\in(0,1)$, set $Z = \frac{9mM}{\eta}(\geq 9)$, $h = d^{-2} \frac{\log\log Z}{2\log Z}$ and $N=Z \log^4 Z = \Otilde(\nicefrac{mM}{\eta})$. Then, the failure probability of one iteration of $\fb$ is at most $\eta/m$, and the expected membership queries per iteration is $\O(M \log^4\nicefrac{mM}{\eta})$.
\end{lemma}

{
The bias of conditioning on success can be bounded as follows.
\begin{lemma}[Bias from no failure]\label{lem:bias-from-failure}
Let $\msf{Succ}$ be the event that the failure is not declared over the required number $m$ iterations, and suppose $\P(\msf{Succ})\geq 1 - \eta$. Let $\mu_m$ be the law of the uncapped $\fb$ output after $m$ iterations, and $\mu_m^{\textup{cap}}$ be the law of Algorithm~\ref{alg:fb-scheme-unif} conditioned on $\msf{Succ}$. Then, we have:
\begin{align*}
    \frac{\D \mu_m^{\textup{cap}}}{\D \mu_m} &\leq \frac{1}{1-\eta}     \quad \textup{so}\quad \eu{R}_\infty(\mu_m^{\textup{cap}} \| \mu_m) \leq \log\frac{1}{1-\eta}\,,\\
\eu{R}_q(\mu_m^{\textup{cap}} \,\|\, \pi^X) & \leq \eu{R}_q(\mu_m \|\pi) + \frac{q}{q-1}\,\log\frac{1}{1-\eta}\qquad\text{for }q>1\,.
\end{align*}
\end{lemma}
}

\subsection{Functional inequalities}\label{sec:func-ineq}

The contraction of an outer loop of our algorithm is controlled by isoperimetry of the uniform distribution $\pi^X$, which is described precisely by a functional inequality. The most natural ones to consider in this setting are the Poincar\'e inequality~\eqref{eq:pi} and log-Sobolev inequality~\eqref{eq:lsi}. In \S\ref{app:ftn-ineq}, we provide a more detailed discussion of how these are related to other important notions of isoperimetry, such as the \emph{Cheeger} and \emph{log-Cheeger} inequalities. 

Below, we use $\mu, \nu$ to denote two arbitrary probability measures over $\Rd$.
The relationship between a Poincar\'e inequality and the $\chi^2$-divergence is derived by substituting $f = \frac{\D \nu}{\D \mu}$ into~\eqref{eq:pi}.
\begin{lemma}\label{lem:pi-to-chi}
    Assume that $\nu$ satisfies \eqref{eq:pi} with parameter $C_{\msf{PI}}(\nu)$. For any probability measure $\mu$ over $\Rd$ with $\mu \ll \nu$, it holds that
    \[
    \chi^2(\mu \mmid \nu) \leq \frac{C_{\msf{PI}}(\nu)}{2}\,\E_\nu\bbrack{\big\lvert\nabla \frac{\D \mu}{\D \nu}\big\rvert^2}\,.
    \]
\end{lemma}

The Poincar\'e inequality implies functional inequalities for the R\'enyi divergence.

\begin{lemma}[{\cite[Lemma 9]{vempala2019rapid}}]\label{lem:pi-renyi}
    Assume that $\nu$ satisfies \eqref{eq:pi} with parameter $C_{\msf{PI}}(\nu)$. For any $q\geq 2$ and probability measure $\mu$ over $\Rd$, it holds that
    \[
    1-\exp\bigl(-\eu R_q(\mu \mmid \nu)\bigr) \leq \frac{q\,C_{\msf{PI}}(\nu)}{4}\, \msf{RF}_{q}(\mu \mmid \nu)\,,
    \]
    where $\msf{RF}_{q}(\mu \mmid \nu):= q\, \E_\nu[(\frac{\D\mu}{\D \nu})^{q}\,\abs{\nabla \log\frac{\D\mu}{\D\nu}}^2] \,/\, \E_{\nu}[(\frac{\D\mu}{\D \nu})^{q}] $ is the \emph{R\'enyi Fisher information} of order $q$ of $\mu$ with respect to $\nu$.
\end{lemma}

The log-Sobolev inequality paired with the KL-divergence~\eqref{eq:lsi2} can be understood as a special case of the following inequality\footnote{Such inequalities are often called Polyak-\L ojasiewicz inequalities, which say for $f: \R^d \to \R$, and all $y \in \R^d$ that $f(y) \leq c\,\abs{\nabla f(y)}^2$ for some constant $c$, if $\min f(x) = 0$.} paired with the $q$-R\'enyi divergence for $q\geq1$.

\begin{lemma}[{\cite[Lemma 5]{vempala2019rapid}}]\label{lem:Renyi-LSI}
    Assume that $\nu$ satisfies \eqref{eq:lsi2} with parameter $C_{\msf{LSI}}(\nu)$. For any $q\geq 1$ and probability measure $\mu$ over $\Rd$, it holds that
    \[
    \eu R_q(\mu \mmid \nu) \leq \frac{q\,C_{\msf{LSI}}(\nu)}{2}\, \msf{RF}_{q}(\mu \mmid \nu)\,.
    \]
    Note that $\lim_{q\to 1}\eu R_q = \KL$ and $\msf{RF}_1 = \FI$.
\end{lemma}

We have collected below the functional inequalities used to establish the mixing of our algorithm (see \S\ref{app:ftn-ineq} for a detailed presentation).

\begin{lemma}\label{lem:func-ineq-coll}
    Let $\mc K\subset \Rd$ be a convex body with diameter $D$, and $\pi$ be the uniform distribution over $\mc K$. Then, $C_{\msf{PI}}(\pi)\lesssim \norm{\cov \pi}_{\msf{op}}\, \log d$ and $C_{\msf{LSI}}(\pi) \lesssim D^2$. If $\pi$ is isotropic, then $C_{\msf{PI}}(\pi)\lesssim \log d$ and $C_{\msf{LSI}}(\pi) \lesssim D$.
\end{lemma}

\subsection{Contraction and mixing}\label{sec:contraction-analysis}
We start by analyzing how many outer iterations of $\fb$ are required to be $\veps$-close to $\pi^X$, the uniform distribution over $\mc K$.
The contraction of Algorithm~\ref{alg:fb-scheme-unif} comes from analyzing Lines~\ref{line:forward} and~\ref{line:implement-back} through the perspective of heat flows (see \S\ref{sec:alg-stoch}).
Unlike prior works on uniform sampling, we do not bound the $s$-conductance of this chain. Instead, we use the ``calculus'' of the space of probability measures; i.e., for some probability divergence $D$, we compute $\partial_t D(\mu * \gamma_t \mmid \pi^X * \gamma_t)$ in terms of $t$ and isoperimetric constants of $\pi^X$.

The classical data processing inequality (DPI; see Lemma~\ref{lem:DPI}) shows that $f$-divergence between two distributions cannot increase when both are convolved with a Gaussian. 
To prove contraction, we will use a {\em strong data-processing inequality} (SDPI) for $f$-divergences, which gives a quantitative bound on the contraction. Such an inequality was known for some cases including the KL-divergence~\cite{stam59inequalities}, and was proven for $f(x)=x^q-1$ by Klartag and Ordentlich~\cite{KO25strong} who also established the SDPI more generally for this type of Gaussian convolution \cite{AG76spreading,PW16dissipation,CPW18strong}, showing that finite fourth moment (i.e., $\E_{\pi^X}[\abs{\cdot}^4]<\infty$) is a sufficient condition. 
In \S\ref{app:contraction-sketch}, we show how the identity is derived under suitable regularity assumptions (see Lemma~\ref{lem:f-diff-ineq}).

\begin{proposition}[General de Bruijn identity, \cite{KO25strong}]\label{prop:debruijn}
Let $\mu$ and $\nu$ be probability measures on $\Rd$ such that $\E_\nu[\abs{\cdot}^4]<\infty$ and $\chi^q(\mu \mmid \nu) < \infty$. Then, for any $t>0$ and $q > 1$,
\[
\partial_t \chi^q(\mu_t \mmid \nu_t)
= - \frac{q(q-1)}2\,\E_{\nu_t}\bbrack{\bpar{\frac{\D \mu_t}{\D \nu_t}}^q \bigl|\nabla \log \frac{\D \mu_t}{\D \nu_t}\bigr|^2}
= - \frac{q-1}2\,\Bnorm{\frac{\D \mu_t}{\D \nu_t}}^q_{L^q(\nu_t)} \msf{RF}_q(\mu_t \mmid \nu_t)\,.
\]
\end{proposition}

Using the general de Bruijn identity with $\eu R_q = \frac{1}{q-1}\log(1+\chi^q)$, the chain rule results in
\begin{equation}
\partial_t \eu R_q (\mu_t \mmid \nu_t) 
= \frac{1}{q-1}\, \frac{\partial_t \chi^q(\mu_t \mmid \nu_t)}{1+\chi^q(\mu_t \mmid \nu_t)}
= \frac{1}{q-1}\, \frac{\partial_t \chi^q(\mu_t \mmid \nu_t)}{\norm{\frac{\D \mu_t }{\D \nu_t}}^q_{L^q(\nu_t)}}
= -\frac{1}2\, \msf{RF}_q(\mu_t \mmid \nu_t)\,.\label{eq:renyi-deriv}    
\end{equation}
We also obtain $\partial_t \chi^2(\mu_t \mmid \nu_t) = -\E_{\nu_t}[\abs{\nabla \frac{\D \mu_t}{\D \nu_t}}^2]$.

Before proceeding, we need a property of the stability of functional inequalities under the heat flow.

\begin{lemma}[{Functional inequalities under Gaussian convolutions, \cite[Corollary 13]{chafai2004entropies}}]\label{lem:isoperimetry_heat}
    The following inequality holds for any $t>0$ and $\pi$ with finite log-Sobolev and Poincar\'e constants,
    \[
        C_{\msf{PI}}(\pi_t) \leq C_{\msf{PI}}(\pi) + t\,, \qquad\text{and}\qquad C_{\msf{LSI}}(\pi_t) \leq C_{\msf{LSI}}(\pi) + t\,.
    \]
\end{lemma}

We record contraction results under the heat flow {and functional inequalities in $q$-R\'enyi divergence, combining \cite{chen2022improved} and Proposition~\ref{prop:debruijn}}.
\begin{lemma}[Contraction under functional inequalities]\label{lem:contraction-fi}
    Let $\mu, \nu$ be probability measures on $\Rd$ such that $\eu{R}_q(\mu\mmid \nu) <\infty$ {and $\E_\nu[\abs{\cdot}^4] < \infty$}. Then, for any $t > 0$ and $q > 1$,
    \[
    \eu R_q(\mu_t \mmid \nu_t) \leq \frac{\eu R_q(\mu  \mmid \nu )}{(1+t/C_{\msf{LSI}}(\nu))^{1/q}}\,.
    \]
    Also, $\chi^2(\mu_t \mmid \nu_t) \leq \frac{\chi^2(\mu \mmid \nu)}{(1+t/C_{\msf{PI}}(\nu))}$.
    Furthermore, for any $q \geq 2$,
    \[
    \eu R_{q}(\mu_t \mmid \nu_t)\leq
    \begin{cases}
    \eu R_{q}(\mu \mmid \nu) - \frac{\log(1 + t/C_{\msf{PI}}(\nu))}{q} & \text{if }\eu R_q(\mu \mmid \nu)\geq 1\,,\\
    \frac{\eu R_q(\mu \mmid \nu)}{(1 + t/C_{\msf{PI}}(\nu))^{1/q}} & \text{if }\eu R_q(\mu \mmid \nu)<1\,.
    \end{cases}
    \]
\end{lemma}

\begin{proof}
    It is well-known that \eqref{eq:lsi} implies \eqref{eq:pi}, and that \eqref{eq:pi} ensures finite moments of any order due to exponential integrability (see \cite[Proposition 4.4.2]{BGL14analysis}). Also, both $\mu_t$ and $\nu_t$ are smooth due to the Gaussian convolution.

    Under $C_{\msf{LSI}}(\nu) < \infty$, \eqref{eq:renyi-deriv} implies that
    \[
    \partial_t \eu R_q(\mu_t \mmid \nu_t)
     = - \half\, \msf{RF}_q(\mu_t \mmid \nu_t)
    \underset{(i)}{\leq} - \frac{\eu R_q(\mu_t \mmid \nu_t)}{q\,C_{\msf{LSI}}(\nu_t)}
    \underset{(ii)}{\leq} -\frac{1}q\, \frac{\eu R_q(\mu_t \mmid \nu_t)}{C_{\msf{LSI}}(\nu) + t}\,,
    \]
    where we used Lemma~\ref{lem:Renyi-LSI} in $(i)$ and Lemma~\ref{lem:isoperimetry_heat} in $(ii)$.
    Applying Gr\"onwall's inequality (Lemma~\ref{lem:gronwall}), we deduce that
    \[
    \eu R_q(\mu_t \mmid \nu_t)
    \leq  \exp \Bpar{-\frac{1}{q} \int_0^t \frac{1}{C_{\msf{LSI}}(\nu) + s}\, \D s}\, \eu R_q(\mu \mmid \nu) \leq \frac{\eu R_q(\mu  \mmid \nu )}{(1+t/C_{\msf{LSI}}(\nu))^{1/q}}\,.
    \]

    The result in the $\chi^2$-divergence can be derived entirely analogously. For instance, the decay from the forward part can be shown as follows:
    \[
        \partial_t \chi^2(\mu_t \mmid \nu_t) 
        = -\frac{1}{2}\, \E_{\nu_t} \bbrack{\big\lvert\nabla \frac{\mu_t}{\nu_t}\big\rvert^2}
        \underset{(i)}{\leq} - \frac{\chi^2(\mu_t \mmid \nu_t)}{C_{\msf{PI}}(\nu_t)}
        \leq - \frac{\chi^2(\mu_t \mmid \nu_t)}{C_{\msf{PI}}(\nu) + t} \,,
    \]
    where $(i)$ follows from Lemma~\ref{lem:pi-to-chi}. Applying Gr\"onwall's inequality then gives
    \[
        \chi^2(\mu_t \mmid \nu_t) \leq \exp \Bpar{-\int_0^t \frac{1}{C_{\msf{PI}}(\nu) + s}\, \D s}\, \chi^2(\mu \mmid \nu)
        \leq \frac{\chi^2(\mu \mmid \nu)}{1 + t/C_{\msf{PI}}(\nu)}\,.
    \]
    
    The result in the $\eu R_q$ under \eqref{eq:pi} can be shown in a similar manner. The only difference is that in forward computation, one should use the functional inequality in Lemma~\ref{lem:pi-renyi} and the following standard inequalities:
    \[
    1-\exp\bpar{-\eu R_{q}(\mu\mmid\nu)}\geq
    \begin{cases}
    \half & \text{if }\eu R_{q}(\mu\mmid\nu)\geq1\,,\\
    \half\eu R_{q}(\mu\mmid\nu) & \text{if }\eu R_{q}(\mu\mmid\nu)\leq1\,.
    \end{cases}
    \]
    This completes the proof.
\end{proof}

Using the contraction result above, we conclude the mixing time of $\fb$.

\begin{proof}[Proof of Theorem~\ref{thm:cont-uni-dist}]
    Note that one iteration of $\fb$ corresponds to the forward and backward step. 
    The forward step \eqref{eq:forward-heat} (i.e., transition kernel $P_h = \mc N(x,hI_d)$) convolves the law $\mu_k^X$ of the $k$-th iterate with Gaussian $\gamma_h$, so the forward step invokes the contraction with rate dependent on $C_{\msf{PI}}(\pi^X)$ or $C_{\msf{LSI}}(\pi^X)$ by Lemma~\ref{lem:contraction-fi}.
    Using the DPI (Lemma~\ref{lem:DPI}) to address the effect of the backward step \eqref{eq:backward-heat}, we obtain that
    \[
    \eu R_q(\mu_{k+1}^X \mmid \pi^X) 
    \underset{\text{DPI}}{\leq} \eu R_q(\mu_{k+1}^Y \mmid \pi^Y)
    \leq \frac{\eu R_q(\mu_k^X \mmid \pi^X)}{(1+h/C_{\msf{LSI}}(\pi^X))^{1/q}}\,.
    \]
    Induction on $k$ proves the first claim. The other results can be proven in the similar fashion.
\end{proof}

\subsection{Failure probability and wasted steps}\label{sec:fail-prob}

We begin by defining a Gaussian version of \emph{local conductance} \cite{kannan1997random}.

\begin{definition}[Local conductance]\label{def:local-conductance}
The local conductance $\ell$ on $\Rd$ is defined by 
\[
\ell(x) \defeq \frac{\int_{\mc K}\exp(-\frac{1}{2h}\abs{x-y}^{2})\,\D y}{\int_{\Rd}\exp(-\frac{1}{2h}\abs{x-y}^{2})\,\D y}
= \frac{\int_{\mc K}\exp(-\frac{1}{2h}\abs{x-y}^{2})\,\D y}{(2\pi h)^{d/2}}\,.
\]
\end{definition}
The local conductance at $y$ quantifies the success probability of the proposal at $y$ in Line~\ref{line:implement-back}.
Then the expected number of trials until the first success of Line~\ref{line:implement-back} is $1/\ell(y)$. Revisiting \eqref{eq:unif-conv}, we can notice $\pi^Y(y) = \ell(y) / \vol(\mc K)$.

\para{Na\"ive analysis for expected number of trials.}
Starting from $\pi^X$, when we just na\"ively sample from $\pi^{Y|X}(\cdot|x)$ for all $x$ without imposing any \emph{failure} condition, the expected number of trials for one iteration is that for the probability density $p_x$ of $\mc N(x,hI_d)$,
\[
\int_{\mc K}\int_{\Rd}\frac{1}{\ell(y)}\,p_{x}(\D y)\pi^X(\D x)  
= \int_{\Rd}\frac{1}{\ell(y)}\,\pi^Y(\D y)
= \int_{\Rd}\frac{1}{\ell(y)}\,\frac{\ell(y)}{\vol(\mc K)} \,\D y=\infty\,.
\]
This suggests that one should consider the algorithm as having ``failed'' if the number of trials exceeds some threshold. 

\para{Refined analysis under a failure condition.}
Going forward, we assume an $M$-warm start as in previous work for uniform sampling algorithms. By induction we have $\frac{\D\mu^X_{i}}{\D\pi^X}\leq M$ for all $i$.

\begin{lemma}[Propagation of warm-start]
From an $M$-warm start, we have $\nicefrac{\D\mu_{i}^{X}}{\D\pi^X} \leq M$ for all $i$.    
\end{lemma}

\begin{proof}
Assume that $\mu_{i}^{X}$ satisfies the $M$-warm start. Then, for any measurable $S$ and the transition kernel $T_{x}$ of Algorithm~\ref{alg:fb-scheme-unif} at $x$,
\[
\mu_{i+1}^{X}(S)=\int_{\mc K}T_{x}(S)\,\D\mu_{i}^{X}(x)\leq M\int_{\mc K}T_{x}(S)\,\D\pi^X(x)=M\pi^X(S)\,,
\]
where the last equality follows from the stationarity of $\pi$. Hence, $\D\mu_{i+1}^{X}/\D\pi^X\leq M$.     
\end{proof}

We now establish a lemma that comes in handy when analyzing the failure probability of the algorithm. In essence, this lemma bounds the probability that taking a Gaussian step from $\pi^X$ in Line~\ref{line:forward} gets $\delta$-distance away from $\mc K$. Let us denote the $\delta$-blowup of $\mc K$ by $\mc K_{\delta} := \{x\in \Rd: d(x, \mc K)\leq \delta\}$.

\begin{lemma}
For a convex body $\mc K\subset \Rd$ containing a unit ball $B_1(0)$,
\[
\pi^Y(\mc K_\delta^c) \leq \exp\bpar{-\frac{\delta^2}{2h}+\delta d}\,.
\]
\end{lemma}

\begin{proof}
For $y\in\de \mc K_{\delta}$, we can take the supporting half-space $H(y)$ at $\msf{proj}_{\mc K}(y)$ containing $\mc K$, due to convexity of $\mc K$. Then,
\begin{align}
\pi^{Y}(\mc K_{\delta}^{c})
&=\frac{1}{\vol(\mc K)}\int_{\mc K_{\delta}^{c}}\int_{\mc K}\frac{\exp(-\frac{1}{2h}\abs{y-x}^2)}{(2\pi h)^{d/2}}\,\D x\, \D y \nonumber
\\
&\le
\frac{1}{\vol(\mc K)} \int_{\mc K_{\delta}^{c}}\int_{H(y)}\frac{\exp(-\frac{1}{2h}\abs{y-x}^2)}{(2\pi h)^{d/2}}\,\D x \, \D y \nonumber\\
&=
\frac{1}{\vol(\mc K)} \int_{\mc K_{\delta}^{c}} 
\int_{d(y,\mc K)}^{\infty} \frac{\exp(-\frac{z^2}{2h})}{\sqrt{2\pi h}}\,\D z \, \D y\,.\label{eq:piy-bound}
\end{align}

Let us denote the tail probability of the $1$-dimensional Gaussian with variance $h$ by
\[
\msf T(s) := \P_{\mc N(0,h)}(Z\geq s)
= 1 - \Phi(h^{-1/2}s)\,,
\]
where $\Phi$ is the CDF of the standard Gaussian.
By the co-area formula and integration by parts,
\begin{align}
&\int_{\mc K_{\delta}^{c}} \int_{d(y,\mc K)}^{\infty} \frac{\exp(-\frac{1}{2h}z^2)}{\sqrt{2\pi h}}\,\D z \D y 
 =\int_{\delta}^{\infty} \msf T(s) \vol(\de \mc K_{s})\,\D s\nonumber\\
& =\Bbrack{\underbrace{\msf T(s)\int_{0}^{s}\vol(\de \mc K_{z})\,\D z}_{\eqqcolon \msf F}}_{\delta}^\infty + \int_{\delta}^{\infty}\frac{1}{\sqrt{2\pi h}}\exp\bpar{-\frac{s^{2}}{2h}}\int_{0}^{s}\vol(\de \mc K_{z})\,\D z\,\D s\,. \label{eq:coarea-ibp}
\end{align}
Recall that $\msf T(s)\leq\half\exp(-\half(h^{-1/2}s)^{2})$ for $h^{-1/2} s \geq 0$ due to a standard tail bound on a Gaussian distribution. 
This tail bound, combined with
\[
\int_{0}^{s}\vol(\de \mc K_{z})\,\D z 
=\vol(\mc K_{s})-\vol(\mc K)\leq\vol\bpar{(1+s)\, \mc K}-\vol(\mc K)
=\bpar{(1+s)^{d}-1}\vol(\mc K)\,,
\]
ensures that $\msf F$ vanishes at $s=\infty$. Hence, bounding the first term in \eqref{eq:coarea-ibp} by $0$ results in
\begin{align*}
\int_{\mc K_{\delta}^{c}} \int_{d(y,\mc K)}^{\infty} \frac{\exp(-\frac{1}{2h}z^2)}{\sqrt{2\pi h}}\,\D z\, \D y
& \leq\frac{1}{\sqrt{2\pi h}}\int_{\delta}^{\infty}\exp\bpar{-\frac{s^{2}}{2h}}\bpar{\underbrace{(1+s)^{d}}_{\leq \exp(sd)}-1}\vol(\mc K)\,\D s\\
& \leq \frac{\vol(\mc K)}{\sqrt{2\pi h}}\exp(hd^{2}/2)\int_{\delta}^{\infty}\exp\bpar{-\frac{1}{2h}\,(s-hd)^{2}}\,\D s\\
& \underset{(i)}{\leq}\vol(\mc K)\exp(hd^{2}/2)\exp\bpar{-\frac{(\delta - hd)^2}{2h}}\\
& = \vol(\mc K)\exp\bpar{-\frac{\delta^2}{2h} + \delta d}\,,
\end{align*}
where in $(i)$ we used the tail bound for a Gaussian.
\end{proof}

This core lemma suggests taking $\delta = \nicefrac{t}d$ and $h=\nicefrac{c}{d^2}$ for some $t,c>0$, under which we have
\[
\pi^Y(\mc K_\delta^c) \leq \exp\bpar{-\frac{t^2}{2c}+t}\,.
\]
Now we choose a suitable threshold $N$ for bounding the failure probability. Following \eqref{eq:piy-bound} in the proof, one can notice that for $y\in \mc K^c_\delta$, $\delta = \Omega(t/d)$, and $h=\Theta(d^{-2})$,
\[
\ell(y) \leq \int_{d(y,\mc K)}^\infty \frac{\exp(-\frac{1}{2h}z^2)}{\sqrt{2\pi h}} \,\D z = \P_{\mc N(0,h)}(Z\geq \delta) \leq \exp(-\Omega(t^{2}))\,.
\]
Thus, the expected number of trials from $\mc K^c_\delta$ for the rejection sampling in Line~\ref{line:implement-back} is $\ell(y)^{-1} \geq \exp(\Omega(t^2))$.
Intuitively, one can ignore whatever happens in $\mc K^c_{\delta}$, since $\mc K_\delta$ takes up most of measure of $\pi^Y$.
As the number of trials from $\mc K_\delta^c$ is at least $\exp(\Omega(t^2))$ in expectation, the most straightforward way to ignore algorithmic behaviors from $\mc K_\delta^c$ is simply to set the threshold to $N = \Otilde(\exp(t^{2}))$. Even though the threshold is $N$, the expected number of trials is much lower.

Lemma~\ref{lem:per-iteration-guarantees} bounds the failure probability and expected number of trials per iteration.

\begin{proof}[Proof of Lemma~\ref{lem:per-iteration-guarantees}]
For $\mu_h := \mu * \mc N(0,hI_d)$, the failure probability is $\E_{\mu_h}[(1-\ell)^N]$. Since $\D\mu/\D\pi^X\leq M$ implies $\D \mu_h / \D (\pi^X)_h = \D \mu_h / \D \pi^Y\leq M$ (as easily checked by the definition of convolution), it follows that 
\[
\E_{\mu_h}[(1-\ell)^N] \leq M\,\E_{\pi^Y}[(1-\ell)^N]\,. 
\]
Then,
\begin{align*}
\int_{\Rd} \underbrace{(1-\ell)^{N}  \,\D\pi^Y}_{\eqqcolon \msf{A}} & =\int_{\mc K_{\delta}^{c}} \msf{A} + \int_{\mc K_{\delta}\cap[\ell\geq N^{-1}\log(3mM/\eta)]} \msf{A} + \int_{\mc K_{\delta}\cap[\ell<N^{-1}\log(3mM/\eta)]} \msf{A}\\
 & \leq\pi^Y(\mc K_{\delta}^{c})+\int_{[\ell\geq N^{-1}\log(3mM/\eta)]}\exp(-\ell N)\,\D\pi^Y \\
 &\qquad \qquad + \int_{\mc K_{\delta}\cap[\ell<N^{-1}\log(3mM/\eta)]}\frac{\ell(y)}{\vol(\mc K)}\,\D y\\
 & \leq \exp\bpar{-\frac{t^2}{2c}+t}+\frac{\eta}{3mM}+\frac{\log(3mM/\eta)}{N}\,\frac{\vol(\mc K_{\delta})}{\vol(\mc K)}\\
 & \leq \exp\bpar{-\frac{t^2}{2c}+t}+\frac{\eta}{3mM}+\frac{e^t}{N}\,\log\frac{3mM}{\eta}\,,
\end{align*}
where we used $\vol(\mc K_{\delta})\subset\vol((1+\delta)\,\mc K)
=(1+\delta)^{d}\vol(\mc K) \leq e^t \vol(\mc K)$. 
Taking $c = \frac{\log \log Z}{ 2 \log Z}$, $t = \sqrt{8}\log\log Z$, and $N = Z\,(\log Z)^4$, we can bound the last line by $\frac{\eta}{mM}$. Therefore,
\[
\E_{\mu_h}[(1-\ell)^N] \leq M\,\E_{\pi^Y}[(1-\ell)^N]\leq \frac{\eta}{m}\,.
\]

We now bound the expected number of trials per iteration. Let $X$ be the minimum of the threshold $N$ and the number of trials until the first success. Then the expected number of trials per step is bounded by $M\E_{\pi^Y}X$ since $\D\mu_{h}/\D\pi^Y\leq M$.
Thus, 
\begin{align*}
\int_{\Rd}\bpar{\frac{1}{\ell}\wedge N}\,\D\pi^Y & \leq\int_{\mc K_{\delta}}\frac{1}{\ell}\,\D\pi^Y + N\pi^Y(\mc K_{\delta}^{c})=\frac{\vol(\mc K_{\delta})}{\vol(\mc K)} + N\pi^Y(\mc K_{\delta}^{c})\\
& \leq e^t + N\exp\bpar{-\frac{t^2}{2c}+t}
\leq \log^3 Z  + 3 \log^4 Z  =\O\bpar{\log^4\frac{mM}{\eta}}\,.
\end{align*}
Therefore, the expected number of trials per step is $\O(M\log^4\frac{mM}{\eta})$, and the claim follows since each trial uses one query to the membership oracle of $\mc K$.
\end{proof}

{Lastly, we prove a simple lemma showing that the bias due to conditioning on success is bounded in terms of the failure probability $\eta$.
\begin{proof}[Proof of Lemma~\ref{lem:bias-from-failure}]
Let $X$ be the uncapped $\fb$ output after the desired number $m$ of iterations, and $\msf{Succ}$ be the event that the failure is not declared until $m$ iterations. Let $\mu_m^X$ be the law of $X$ and $\mu_m^{\textup{cap}}$ be the law of $X$ conditioned on $\msf{Succ}$. Then, by Bayes' rule,
\[
\frac{\D \mu_m^{\textup{cap}}}{\D \mu_m} (x) = \frac{\P(X=x\,|\, \msf{Succ})}{\P(X=x)} = 
\frac{\P(\msf{Succ}\,|\, X=x)\,\P(X=x)}{\P(\msf{Succ})\,\P(X=x)} \leq \frac{1}{\P(\msf{Succ})} \leq \frac{1}{1-\eta}\,.
\]
Lastly, using the weak triangle inequality for R\'enyi divergence,
\[
\eu{R}_q(\mu_m^{\textup{cap}} \,\|\, \pi) \leq \frac{q}{q-1}\,\eu{R}_\infty(\mu_m^{\textup{cap}} \,\|\, \mu_m) + \eu{R}_q(\mu_m \,\|\, \pi) \leq \eu{R}_q(\mu_m \,\|\, \pi) + \frac{q}{q-1}\log\frac{1}{1-\eta}\,,
\]
which completes the proof.
\end{proof}
}

\subsection{Putting it together}\label{sec:example-main-result}

We can now show that $\fb$ subsumes previous results on uniform sampling from convex bodies (such as the $\bw$ and $\sw$), providing detailed versions of the main results in \S\ref{sec:results}.

We first establish that the query complexity of $\fb$ matches that of the $\bw$ under stronger divergences. Recall that $2\,\norm{\cdot}_{\msf{TV}}^2\leq \KL \leq \log(1+\chi^2)\leq \chi^2$. 

\begin{theorem}\label{thm:main-result}
For any given $\eta,\varepsilon \in (0,1/2)$, $q\geq 2$, $m\in \mathbb{N}$ defined below and any convex body $\mc K$ given by a well-defined membership oracle, consider $\fb$ (Algorithm~\ref{alg:fb-scheme-unif}) with an $M$-warm initial distribution $\mu_0^X$, step size $h = (2d^2\log\frac{9mM}{\eta})^{-1}$, and threshold $N = \Otilde(\frac{mM}{\eta})$. For $\pi^X$ the uniform distribution over $\mc K$, let $\mu_m^X$ be the law of the $m$-th iterate of $\fb$ conditioned on the success.
\vspace{-3pt}
\begin{itemize}
    \item  With probability $1-\eta$, the algorithm iterates this many times without failure, using $\Otilde(qM d^2\, \norm{\cov \pi^X}_{\msf{op}} \log^6\frac{1}{\eta \veps})$ expected number of membership queries in total and achieving 
    $\eu R_q(\mu^X_m \mmid \pi^X)\leq \veps + 4\eta$ after $m = \Otilde(q d^2\,\norm{\cov \pi^X}_{\msf{op}} \log^2\frac{M}{\eta \veps})$ iterations.
    \item For isotropic $\pi^X$,  with probability $1-\eta$, the algorithm achieves $\eu R_q(\mu^X_m \mmid \pi^X)\leq \veps + 4\eta$ with $m = \Otilde(qd^2\log^2\frac{M}{\eta \veps})$ iterations, using $\Otilde(qM d^2 \log^6\frac{1}{\eta \veps})$ membership queries in expectation.
\end{itemize}
\end{theorem}

\begin{proof}
    We just put together Lemma~\ref{lem:per-iteration-guarantees} and Theorem~\ref{thm:cont-uni-dist}.
    For target accuracy $\veps > 0$, we use the $\eu R_q$-decay under \eqref{eq:pi} for $q\geq 2$ in Theorem~\ref{thm:cont-uni-dist}. 
    The $M$-warm start assumption guarantees $\eu R_q(\mu_0^X \mmid \pi^X) \lesssim \log M$. 
    Due to $C_{\msf{PI}}(\pi^X) = \O(\norm{\cov \pi^X}_{\msf{op}}\,\log d)$ (Lemma~\ref{lem:func-ineq-coll}), $\fb$ ``without the cap'' can achieve $\eu R_q(\mu^X_m \mmid \pi^X)\leq \veps$ if it iterates at least
    \[
    m\gtrsim qd^2\,\norm{\cov \pi^X}_{\msf{op}}\log d \log\frac{M}{\eta\veps} \log \frac{mM}{\eta}\quad \text{times.}
    \]
    Note that $x \geq A \log Bx$ for $A, B \geq 1$ is satisfied when $x\gtrsim A\log AB = \Otilde(A\log B)$. Using this, we set 
    \[
    m = \Otilde\bpar{qd^2\,\norm{\cov \pi^X}_{\msf{op}} \log\frac{M}{{\eta\veps}}\log\frac{M}{\eta}}
    = \Otilde\bpar{qd^2\,\norm{\cov \pi^X}_{\msf{op}} \log^2\frac{M}{{\eta\veps}}}\,.
    \]
    
    Since each iteration has $\eta/m$-failure probability by Lemma~\ref{lem:per-iteration-guarantees}, the union bound ensures that the total failure probability is at most $\eta$ throughout $m$ iterations.
    Lastly, each iteration requires $\Otilde(M \log^4\nicefrac{1}{\eta\veps})$ membership queries in expectation by Lemma~\ref{lem:per-iteration-guarantees}.
    Therefore, $\fb$ uses $\Otilde(qM d^2\,\norm{\cov \pi^X}_{\msf{op}} \log^6\nicefrac{1}{\eta \veps})$ expected number of membership queries over $m$ iterations.
    Since $\eu R_q$ is non-decreasing in $q$, we can obtain the desired bound on $\eu R_q$ for $q\in [1,2)$.

    For isotropic $\pi^X$, we have $\cov \pi^X = I_d$, so the claim immediately follows from $C_{\msf{PI}}(\pi^X) = \O(\log d)$. {Lastly, as noted in Lemma~\ref{lem:bias-from-failure}, the conditional-on-success output distribution of  $\fb$ introduces a small bias, bounded by $2\eta$.}
\end{proof}

{
We show that the restart variant increases the expected total query complexity by at most a factor of $(1-\eta)^{-1}$.
\begin{proof}[Proof of Corollary~\ref{cor:comp-restart}]
Since the success probability of $m$ iterations without failure is at least $1-\eta$, the expected number of restarts is $(1-\eta)^{-1}$. Setting $4\eta = \veps (< 1/10)$ and replacing $\veps$ by $\veps/2$ in Theorem~\ref{thm:main-result}, the conclusion follows.
\end{proof}
}

We now bound the number of proper steps for general \emph{non-convex bodies} and \emph{any feasible start} in convex $\mc K$. {We first record the following under an $M$-warm start, as a corollary of the contraction results in \cite{chen2022improved, KO25strong}.}

\begin{proposition}\label{thm:conv-lsi-pi}
For any given $\varepsilon\in(0,1)$ and set $\mc K \subset B_D(0)$ with $\vol(\mc K) > 0$, $\fb$ with variance $h$ and $M$-warm initial distribution achieves $\eu R_q(\mu^X_m \mmid \pi^X)\leq \veps$ after the following number of iterations:
    \[
    m= \min 
    \begin{cases}
        \O\bpar{qh^{-1} C_{\msf{PI}}(\pi^{X}) \log\frac{M}{\veps}} & \text{for }q\geq 2\,,\\
        \O\bigl(qh^{-1} C_{\msf{LSI}}(\pi^{X}) \log\frac{\log M}{\veps}\bigr) & \text{for } q \ge 1\,.
    \end{cases}
    \]
\end{proposition}

\begin{proof}
    By the R\'enyi-decay under \eqref{eq:lsi} in Theorem~\ref{thm:cont-uni-dist}, $\fb$ can achieve $\veps$-distance to $\pi^X$ after $\O\bpar{qh^{-1}C_{\msf{LSI}}(\pi^X)\log\frac{\eu R_q(\mu^X_1 \mmid \pi^X)}{\veps}}$ iterations for $q\geq 1$. 
    
    For $q \geq 2$, we use the decay result under \eqref{eq:pi}. In this case, $\fb$ decays under two different rates depending on the value of $\eu R_q(\cdot \mmid \pi^X)$.
    It first needs $\O(qh^{-1}C_{\msf{PI}}(\pi^X)\, \eu R_q(\mu_0^X\mmid \pi^X))$ iterations until $\eu R_q(\cdot \mmid \pi^X)$ reaches $1$. Then, $\fb$ additionally needs $\O(qh^{-1}C_{\msf{PI}}(\pi^X)\log\frac{1}{\veps})$ iterations, and thus it needs $\O(qh^{-1}C_{\msf{PI}}(\pi^X) \bpar{\eu R_q(\mu_0^X \mmid \pi^X)+\log\frac{1}{\veps}})$ iterations in total. By substituting $\eu R_q(\mu_0^X\mmid \pi^X) \lesssim \log M$, we complete the proof.
\end{proof}

Next, we show that $\fb$ mixes from any start.
\begin{proof}[Proof of Corollary~\ref{thm:any-start}]
    We first bound the warmness of $\mu^X_1$ w.r.t. $\pi^{X}$ when $\mu^X_0 = \delta_{x_0}$.
    One can readily check that 
    \[
    \mu_{1}^{X}(x)=\ind_{\mc K}(x) \int\frac{\exp(-\frac{1}{2h}\,\abs{y-x}^{2})\exp(-\frac{1}{2h}\,\abs{y-x_{0}}^{2})}{(2\pi h)^{d/2}\int_{\mc K}\exp(-\frac{1}{2h}\,\abs{y-x}^{2})\,\D x}\,\D y\,.
    \]
    By Young's inequality, $\abs{y-x}^{2}\leq(\abs y+D)^{2}\leq\frac{3}{2}\,\abs y^{2}+3D^{2}$ for $x\in\mc K$.
    Hence,
    \begin{align*}
        &\int\frac{\exp(-\frac{1}{2h}\,\abs{y-x}^{2})\exp(-\frac{1}{2h}\abs{y-x_{0}}^{2})}{\int_{\mc K}\exp(-\frac{1}{2h}\,\abs{y-x}^{2})\,\D x}\,\D y\\
        \leq& \frac{\exp(2h^{-1}D^{2})}{\vol(\mc K)}\int\exp\Bpar{-\frac{1}{2h}\,(\abs{y-x}^{2}+\abs{y-x_{0}}^{2}-\frac{3}{2}\abs y^{2})}\,\D y\\
	=&\frac{\exp(2h^{-1}D^{2})}{\vol(\mc K)}\int\exp\Bpar{-\frac{1}{2h}\bpar{\half\,\abs{y-2(x+x_{0})}^{2}+(\abs x^{2}+\abs{x_{0}}^{2}-2\abs{x+x_{0}}^{2})}}\,\D y\\
	\leq&\frac{\exp(5h^{-1}D^{2})}{\vol(\mc K)}\int\exp\bpar{-\frac{1}{4h}\,\abs{y-2(x+x_{0})}^{2}}\,\D y\\
	=&\frac{\exp(5h^{-1}D^{2})}{\vol(\mc K)}\,(4\pi h)^{d/2}\,.
    \end{align*}
    Therefore, $M=\esssup \frac{\mu_{1}^{X}}{\pi^{X}}\leq2^{d/2}\exp(5h^{-1}D^{2})$.
    By Theorem~\ref{thm:conv-lsi-pi} under \eqref{eq:lsi}, $\fb$ needs $\Otilde(qh^{-1}C_{\msf{LSI}}(\pi^X)\log\frac{d+D^2/h}{\veps})$ iterations.
\end{proof}

We then obtain the following corollary for a convex body $\mc K$.

\begin{proof}[Proof of Corollary~\ref{cor:any-start-convex}]
    For convex $\mc K$, it follows from Lemma~\ref{lem:func-ineq-coll} that $C_{\msf{LSI}}(\pi^X) = \O(D^2)$ and $C_{\msf{LSI}}(\pi^X) = \O(D)$ for isotropic $\mc K$. The rest of the proof can be completed in a similar way.
\end{proof}

For $h=\tilde{\Theta}(d^{-2})$, $\fb$ requires $\Otilde(qd^2D^2)$ iterations and in particular $\Otilde(qd^2D)$ iteration for isotropic uniform distributions. These bounds match those of the $\sw$ \cite{kannan2006blocking,lee2017eldan} (see Theorem~\ref{thm:sw-results}), albeit with stronger guarantees on the output distribution.

\paragraph{Acknowledgements.} We are deeply grateful to Andre Wibisono and Sinho Chewi for helpful comments and pointers to the literature for Lemma~\ref{lem:f-diff-ineq}. This work was supported in part by NSF award 2106444, 
NSERC through the CGS-D award, and a Simons Investigator award.

\bibliographystyle{alpha}
\bibliography{main}
\newpage
\appendix
\section{\texorpdfstring{$\bw$}{Ball walk} and \texorpdfstring{$\sw$}{Speedy walk}} \label{app:bw-sw}

We restate the previously known guarantees for uniform sampling by the $\bw$ and $\sw$.
Below, let $B_r(x)$ denote the $d$-dimensional ball of radius $r$ centered at $x$.

\begin{algorithm}[h!]
\hspace*{\algorithmicindent} \textbf{Input:} initial distribution $\pi_{0}$, convex body $\mc K\subset\R^{d}$, iterations $T$, step size $\delta > 0$.

\begin{algorithmic}[1]
\caption{$\bw$}\label{alg:ball-walk}
\STATE Sample $x_0 \sim \pi_0$.
\FOR{$i=1,\dotsc,T$}
    \STATE Sample $y\sim \mc \Unif(B_{\delta}(x_{i-1}))$.
    \STATE If $y\in \mc K$, then $x_{i}\gets y$. Else, $x_{i} \gets x_{i-1}$.
\ENDFOR
\end{algorithmic}
\end{algorithm}

The $\bw$ is particularly simple; draw a uniform random point from $B_{\delta}$ around the current point, and go there if the drawn point is inside of $\mc K$ and stay at the current point otherwise. Its stationary distribution can be easily seen to be $\pi \propto \ind_{\mc K}$, the uniform distribution over $\mc K$.

In the literature, there are two approaches to analyzing the convergence rate of this sampler: (i) a direct analysis via the $s$-conductance of the $\bw$ and (ii) an indirect approach which first passes through the $\sw$.

\para{Direct analysis.}
The following $\tv$-guarantee is obtained by lower bounding the $s$-conductance of the $\bw$, which requires a one-step coupling argument and the Cheeger inequality for $\pi$. We refer interested readers to \cite[\S5]{vempala2005geometric}.

\begin{theorem}[Convergence of $\bw$]\label{thm:bw}
For any $\veps \in (0, 1)$ and convex body $\mc K\subset \Rd$ presented by a well-defined membership oracle, let $\pi_{t}$ be the distribution after $t$ steps of the $\bw$ with an $M$-warm initial distribution $\pi_0$.
Then, the $\bw$ with step size $\delta = \Theta(\frac{\veps}{M\sqrt{d}})$ achieves $\norm{\pi_t - \pi}_{\tv}\leq \veps$ for $t\gtrsim d^{2}D^{2}\frac{M^{2}}{\veps^{2}}\log\frac{M}{\veps}$. 
If $\pi$ is isotropic, then the $\bw$ needs $\O(d^{2}\frac{M^{2}}{\veps^{2}}\log d\log\frac{M}{\veps})$ iterations. \end{theorem}

The mixing time of the $\bw$ under this approach has a polynomial dependence on $1/\veps$, rather than a polylogarithmic dependence.

\para{Indirect analysis through $\sw$.}
\cite{kannan1997random} introduced the $\sw$, which could be viewed as a version of the $\bw$ and converges to a \emph{speedy distribution} (see Proposition~\ref{lem:speedy-dist}), which is slightly biased from $\pi$. Then, the $\sw$ is used together with another algorithmic component (rejection sampling) \cite[Algorithm 4.15]{kannan1997random} that converts the speedy distribution to the uniform distribution.
In the literature, the $\bw$ often refers to `$\sw$ combined with the conversion step', rather than a direct implementation of Algorithm~\ref{alg:ball-walk}. Strictly speaking, a mixing guarantee of this combined algorithm should not be referred to as a provable guarantee of the $\bw$.

\begin{algorithm}[h]
\hspace*{\algorithmicindent} \textbf{Input:} initial distribution $\pi_{0}$, convex body $\mc K\subset\R^{d}$, iterations $T$, step size $\delta>0$.

\begin{algorithmic}[1]
\caption{$\sw$}\label{alg:speedy-walk-uniform}
\STATE Sample $x_0 \sim \pi_0$.
\FOR{$i=1,\dotsc,T$}
    \STATE Sample $x_i \sim\Unif\Par{\mc K\cap B_{\delta}(x_{i-1})}$. \label{sw-unif-backward}
\ENDFOR
\end{algorithmic}
\end{algorithm}
 
As opposed to the $\bw$, the $\sw$ \emph{always} takes some step at each iteration. 
However, the problem of sampling from $x_i \sim\Unif\Par{\mc K\cap B_{\delta}(x_{i-1})}$ in Line~\ref{sw-unif-backward} is not straightforward.
This step admits the following implementation based on rejection sampling, via a procedure denoted by $(\ast)$:
\begin{itemize}
    \item Propose $y \sim \Unif(B_{\delta}(x_{i-1}))$. 
    \item Set $x_{i+1} \gets y$ if $y\in \mc K$. Otherwise, repeat the proposal.
\end{itemize}

Each actual step (indexed by $i$) in the $\sw$ is called a \emph{proper step}, and rejected steps during $(\ast)$ are called \emph{improper steps}.
For example, if $x_1,x_1,x_2,x_3,x_3,x_3,x_4,\dotsc$ are the positions produced by the $\bw$, then only proper steps $x_1, x_2, x_3, x_4,\dotsc$ are recorded by the $\sw$.

To describe the theoretical guarantees of the $\sw$, we define the \emph{local conductance} $\ell(x)$ at $x\in \mc K$, which measures the success probability of the rejection sampling scheme in $(\ast)$:
\[
\ell(x) := \frac{\vol(\mc K\cap B_{\delta}(x))}{\vol(B_{\delta}(x))}\,,
\]
and define the \emph{average conductance}:
\[
\lambda := \E_{\pi} \ell = \frac{1}{\vol(\mc K)} \int_{\mc K}\ell(x)\, \D x\,.
\]

\begin{proposition}[\cite{kannan1997random}] \label{lem:speedy-dist}
The stationary distribution $\nu$ of the $\sw$ has density
\[
\nu(x) = \frac{\ell(x) \, \ind_{\mc K} (x)}{\int_{\mc K}\ell(x) \,\D x} \,.
\]
\end{proposition}

The speedy distribution $\nu$ is indeed different from the uniform distribution $\pi$, and this discrepancy is quantified in terms of the average conductance.

\begin{proposition}[{\cite[Page 22]{kannan1997random}}]
$\norm{\nu - \pi}_{\tv}\leq\frac{1-\lambda}{\lambda}$.
\end{proposition}

One can relate the step size $\delta$ to the average conductance.

\begin{proposition}[{Bound on average conductance, \cite[{Corollary 4.5}]{kannan1997random}}] $\lambda\geq1-\frac{\delta\sqrt{d}}{2}$.
\end{proposition}

The best known result for $\sw$'s mixing is due to \cite{kannan2006blocking} devising the \emph{blocking conductance} and using the \emph{log-Cheeger} inequality. 
When $\nu$ is isotropic (i.e., it has covariance proportional to the identity matrix), \cite{lee2017eldan} improves the mixing bound via the \emph{log-Cheeger} constant.

\begin{theorem}[Mixing of $\sw$] \label{thm:sw-results}
For any $\veps \in (0, 1)$ and convex body $\mc K\subset \Rd$ presented by a well-defined membership oracle, let $\nu_{t}$ be the distribution after $t$ proper steps of the $\sw$ started at any feasible point $x_{0} \in \mc K$.
Then, the $\sw$ with step size $\delta=\Theta(d^{-1/2})$ achieves $\norm{\nu_t - \nu}_{\tv}\leq \veps$ for $t\gtrsim (D^{2}+\log(D\sqrt{d}))\,d^{2}\log\frac{1}{\veps}$.
From an $M$-warm start, the expected number of improper steps during $t$ iterations is $\Otilde(tM)$.
When $\nu$ is isotropic, the $\sw$ needs $\mc O(d^{2}D  \log\frac{1}{\veps} \log\log D)$ proper steps to achieve $\veps$-$\tv$ distance to $\nu$.
\end{theorem}

Then, \cite{kannan1997random} uses the following post-processing step to obtain an approximately uniform distribution on $\mc K$, with a provable guarantee.
\vspace{-10pt}
\begin{center}
    $\mc A$: Call the $\sw$ to obtain a sample $X \sim \nu_t$ until $\frac{2d}{2d-1}\,X\in\mc K$. If so, return $\bar{X}= \frac{2d}{2d-1}\,X$.
\end{center}

\begin{proposition}[{\cite[Theorem 4.16]{kannan1997random}}]
    Under the same setting above, assume $\norm{\nu_t - \nu}_{\tv} \leq \veps$ for step size $\delta \leq (8d\log\frac{d}{\veps})^{-1/2}$ and fixed $t \in \mathbb{N}$. 
    For $\bar{\nu} = \law(\bar{X})$ given by $\mc A$, it holds that $\norm{\bar{\nu} - \pi}_{\tv}\leq \veps$, and the expected number of calls on the conversion algorithm is at most $2$.
\end{proposition}

Combining the previous two results, we conclude that the total expected number of membership queries to obtain a sample $\veps$-close to $\pi$ in $\tv$ is $\Otilde(Md^2D^2\log\frac{1}{\veps})$, which now has a poly-logarithmic dependence on $1/\veps$.

\section{Functional inequalities} \label{app:ftn-ineq}
We provide full details on functional inequalities omitted in \S\ref{sec:func-ineq}.
We use $\mu$ and $\mu_{\msf{LC}}$ to denote a probability measure and logconcave probability measure over $\Rd$, respectively.

\para{Cheeger and PI constants.}
The \emph{Cheeger isoperimetric constant} $C_{\msf{Ch}}(\mu)$ measures how large surface area a measurable subset with larger volume has, defined by
\[
C_{\msf{Ch}}(\mu) := \inf_{S\subset\Rd} \frac{\mu^+(S)}{\min(\mu(S),\mu(S^c))}\,,
\]
where the infimum is taken over all measurable subsets $S$, and $\mu^+(S)$ is the Minkowski content of $S$ under $\mu$ defined as, for $S^\varepsilon:=\{x\in X: d(x,S) < \varepsilon \}$,
\[
\mu^+(S) := \liminf_{\varepsilon \to 0} \frac{\mu(S^\varepsilon)-\mu(S)}{\varepsilon}\,.
\]

\cite{cheeger1970lower} established $C_{\msf{PI}}(\mu) \lesssim C_{\msf{Ch}}^{-2}(\mu)$\footnote{The opposite direction $C_{\msf{PI}}(\mu_{\msf{LC}}) \gtrsim C_{\msf{Ch}}^{-2}(\mu_{\msf{LC}})$ also holds for logconcave distributions due to \cite{Ledoux94simple}.}, and then \cite{kannan1995isoperimetric} showed that for  covariance matrix $\Sigma_\mu := \E_{\mu}[(\cdot - \E_\mu X)(\cdot - \E_\mu X)^\T]$,
\begin{equation}\label{eq:PI-uniform}
    C_{\msf{Ch}}(\mu_{\msf{LC}}) \gtrsim \frac{1}{(\E_{\mu_{\msf{LC}}}[\abs{X-\E_{\mu_{\msf{LC}}} X}^2])^{1/2}}
= \frac{1}{(\tr\Sigma_{\mu_{\msf{LC}}})^{1/2}}\,.
\end{equation}
This immediately leads to $C_{\msf{PI}}(\pi)\lesssim (\E_{\pi}[\abs{X-\E_{\pi} X}^2])^{1/2} \leq D^2$ for the uniform distribution $\pi$ over a convex body $\mc K$ with diameter $D>0$.

Kannan et al. proposed the \emph{KLS conjecture} in the same paper, which says that
\[
C_{\msf{Ch}}(\mu_{\msf{LC}}) \gtrsim 
 \frac{1}{\norm{\Sigma_{\mu_{\msf{LC}}}}_{\msf{op}}^{1/2}}\,.
\]
While the original result in \cite{kannan1995isoperimetric} ensures $C_{\msf{Ch}}\gtrsim d^{-1/2}$ for an isotropic logconcave distribution (due to $\Sigma = I_d$), this conjecture indeed claims $C_{\msf{Ch}}\gtrsim 1$ for such case.
Following a line of work \cite{lee2017eldan,chen2021almost,klartag2022bourgain,klartag2023logarithmic}, the current bound is
\[
C_{\msf{Ch}}(\mu_{\msf{LC}}) \gtrsim 
 \frac{(\log d)^{-1/2}}{\norm{\Sigma_{\mu_{\msf{LC}}}}_{\msf{op}}^{1/2}}\,,
\]
which implies that $C_{\msf{PI}}(\pi) \lesssim \log d$ when $\pi$ is isotropic for convex $\mc K$.

\para{Log-Cheeger and LSI constants.}
Just as the Cheeger and PI constants are related above, there are known connections between LSI and \emph{log-Cheeger} constants. The log-Cheeger constant $C_{\msf{logCh}}(\mu)$ of a distribution $\mu \in \mc P(\Rd)$ is defined as
\[
C_{\msf{logCh}}(\mu) := \inf_{S\subset\Rd: \mu(S)\leq \half} \frac{\mu^{+}(S)}{\mu(S)\sqrt{\log\frac{1}{\mu(S)}}}\,.
\]

\cite{Ledoux94simple} established that $C_{\msf{LSI}}(\mu)\lesssim C_{\msf{logCh}}^{-2}(\mu)$ and $C_{\msf{LSI}}(\mu_{\msf{LC}})\gtrsim C_{\msf{logCh}}^{-2}(\mu_{\msf{LC}})$.
\cite{FK99lsi} showed that any logconcave distributions with support of diameter $D>0$ satisfy $C_{\msf{logCh}}(\mu_{\msf{LC}}) \gtrsim D^{-1}$. Later in 2016, \cite{lee2017eldan} improved this to $C_{\msf{logCh}}(\mu_{\msf{LC}}) \gtrsim D^{-1/2}$  under isotropy. 
Therefore, for convex $\mc K$,  it follows that $C_{\msf{LSI}}(\pi) \lesssim D^2$ and that $C_{\msf{LSI}}(\pi) \lesssim D$ if $\pi$ is isotropic.

\section{Semigroup calculus for contraction under the heat flow}\label{app:contraction-sketch}

In this part, we first recall \emph{Markov semigroups}, which are well-established mathematical tools that abstract an underlying Markov process. 
Interested readers can refer to~\cite{BGL14analysis, chewi2023log} for references.
We then review the contraction results for heat flow and its time-reversal~\cite{chen2022improved}, which are intimately connected with our algorithm. We also provide key technical ingredients needed for its proof, such as the computations for measures evolving under simultaneous forward/backward heat flows.

\para{Forward heat flow.} We begin by introducing the ``heat flow'' equation (or also known as the \emph{Fokker--Planck} equation), which describes the evolution of the law of $Z_t$ under \eqref{eq:forward-heat},
\begin{equation}\label{eq:forward-heat2}
    \partial_t \mu_t = \frac{1}{2}\, \Delta \mu_t = \frac{1}{2} \Div(\mu_t \nabla \log \mu_t)\,. \tag{$\msf{FP\text{-}FH}$}
\end{equation}
It is well known that one can realize this equation in discrete time through a Gaussian transition density, in the sense that, for $\mu_h$ (the solution at time $h > 0$ to \eqref{eq:forward-heat2} with initial condition $\mu_0$), and for any smooth function $f: \Rd \to \R$,
\[
    \E_{\mu_h}[f(x)] = \E_{\mu_0}[P_h f(x)]\,,
\]
where $P_h f(x) = \E_{\mc N(x, hI_d)}f(\cdot)$.\footnote{$\{P_h\}_{h \geq 0}$ is often called the heat semigroup.} By this we can formally identify $\mu_h = \mu_0 P_h$, and also write $\mu_h$ for the law of $Z_h$, where $\{Z_h\}_{h \geq 0}$ solves~\eqref{eq:forward-heat}.

\para{Backward heat flow.}
Although there are many ways to define a ``reversal'' of $P_h$, we will use the notion of \emph{adjoint} introduced by~\cite{klartag2021spectral}, which is the most immediately useful.

Given some initial measure $\nu$ and some time horizon $h$, the adjoint corresponds to reversing~\eqref{eq:forward-heat} for times in $[0,h]$ when the initial distribution under consideration is $Z_0\sim \nu$. For other measures, it must be interpreted more carefully, and is given by the following partial differential equation starting from some measure $\mu_0^\leftarrow$ (see  \eqref{eq:awass-sde} and its derivation):
\begin{equation}\label{eq:backward-heat2}
    \partial_t \mu_t^\leftarrow = -\Div\bigl(\mu_t^\leftarrow \nabla \log (\nu P_{h-t})\bigr) + \frac{1}{2}\, \Delta \mu_t^\leftarrow \quad \text{for }t\in[0,h]\,. \tag{$\msf{FP\text{-}BH}$}
\end{equation}
Write $\mu_t^\leftarrow = \mu_0^\leftarrow Q_t^{\nu,h}$, where $\{Q_t^{\nu, h}\}_{t \in [0, h]}$ is a family of transition densities. Write $\mathbf{P}_{0, h}$ for the joint distribution of the $(Z_0, Z_h)$-marginals of~\eqref{eq:forward-heat}, when $Z_0 \sim \nu$, and $\mathbf{P}_{0|h}$ for the conditional. Note that $\mathbf{P}_{h|0}(\cdot|x) = \mc N(x, hI_d)$.
It is also known that~\eqref{eq:backward-heat2} gives a time-reversal of the heat equation at the SDE level, in the sense that we can interpret $\delta_x Q_h^{\nu, h} = \mathbf{P}_{0|h}(\cdot|Z_h = x)$. Thus $\mu_0^\leftarrow Q_h^{\nu, h} = \int \mathbf{P}_{0|h}(\cdot|Z_h = x)\, \mu_0^\leftarrow(\D x)$, and $\nu P_h Q_t^{\nu, h} = \nu P_{h-t}$ for all $t \in [0,h]$.

The ultimate purpose of this machinery is to affirm our earlier description of the Gibbs sampling procedure as alternating forward and backward heat flows. Indeed, notice that, if $\mu_{i}^X$ is the law of the iterate at some iteration $i$, then $\mu_i^X P_h$ is precisely $\mu_{i+1}^Y$ under our scheme, while $(\mu_{i}^X P_h) Q_h^{\pi^X, h}$ is $\mu_{i+1}^X$, assuming $Q_h^{\pi^X, h}$ is well defined for non-smooth measures $\pi^X$. Thus, while Algorithm~\ref{alg:fb-scheme-unif} is implemented via discrete steps, it can be exactly analyzed through arguments in continuous time.

\para{Fokker--Planck equation and time-reversal of SDE.}
The description above can be further generalized further as follows.
Consider a stochastic differential equation $(X_t)$ given by
\begin{equation}\label{eq:awass-sde}
    \D X_t = - a_t(X_t) \, \D t + \, \D B_t \qquad \text{with } X_0 \sim \mu_0\,.
\end{equation}
It is well known that measures $\mu_t$ described by
\begin{equation}\label{eq:awass-Fokker--Planck}
    \partial_t \mu_t = \msf{div}(\mu_t a_t)+ \frac{1}{2} \Delta \mu_t\,,
\end{equation}
correspond to $\law(X_t)$. In this context, \eqref{eq:awass-Fokker--Planck} is referred to as the \emph{Fokker--Planck equation} corresponding to~\eqref{eq:awass-sde}.

From this equation, one can deduce the Fokker--Planck equation of the time reversal $\mu_t^\leftarrow := \mu_{T-t}$:
\[
\de_t \mu_t^\leftarrow = -\Div(\mu_t^\leftarrow a_{T-t}) -\half \Delta \mu_t^\leftarrow
= -\Div\bpar{\mu_t^\leftarrow (a_{T-t} + \nabla \log \mu_{T-t})} +\half \Delta \mu_t^\leftarrow
\]
In particular, this describes the evolution of $\law(X_t)$ of  the stochastic differential equation:
\begin{equation}\label{eq:awass-rev-sde}
    \D X_t = \bpar{a_{T-t}(X_t) + \nabla \log \mu_{T-t}(X_t)} \,\D t +\, \D B_t \qquad \text{with } X_0 \sim \mu_0^\leftarrow = \mu_T\,.
\end{equation}
While the law of this process will give $\mu_T^\leftarrow = \mu_0$ at time $T$, it is also true that it will give $\mu_{0|T}(\cdot\,|\,z)$ if one starts~\eqref{eq:awass-rev-sde} at $X_0 = z$. This is a subtle fact, whose justification requires the introduction of a tool called \emph{Doob's $h$-transform}. The presentation of this subject is beyond the scope of this paper, and we refer interested readers to~\cite{klartag2021spectral} as a reference to its application in this context.

\para{Contraction under simultaneous evolution.}
Instead of considering the change in metrics along the evolution of $\mu P_t$ with respect to ``fixed'' $\nu$, it will be useful to consider the \emph{simultaneous} evolution of $\mu P_t, \nu P_t$ (and similarly $\mu Q_t^{\pi^X, h}, (\nu P_h)Q_t^{\pi^X, h}$). 
This type of computation was carried out for specific metrics in earlier work~\cite{vempala2019rapid,chen2022improved}.
The following is a more generalized form of one appearing in~\cite[Lemma 2]{yuan2023class}. In the lemma below, we consider an arbitrary diffusion equation with corresponding Fokker--Planck equation:
\begin{equation}\label{eq:arbitrary_diffusion}
    \D X_t = b_t(X_t) \, \D t + \, \D B_t\,  
    \quad\text{and}\quad   \partial_t \mu_t = -\nabla \cdot (b_t \mu_t) + \frac{1}{2}\, \Delta \mu_t\,
\end{equation}
where $b_t: \R^d \to \R^d$ is smooth, $X_t \in \R^d$, and $\mu_t = \law(X_t)$ if $X_0 \sim \mu_0$.

Below, we derive contractions of two measures along the same stochastic process, which proves Proposition~\ref{prop:debruijn} under suitable regularity assumptions.

\begin{lemma}[{Decay along forward/backward heat flows}]\label{lem:f-diff-ineq}
    Let $(\mu_t)_{t \geq 0}, (\nu_t)_{t \geq 0}$ denote the laws of the solutions to \eqref{eq:arbitrary_diffusion} starting at $\mu_0, \nu_0$ respectively. 
    Assume that $\mu_t$ and $\nu_t$ has sufficient regularity and fast tail decay so that we can discard boundary terms in integration by parts and differentiate under the integral sign.
    Then, for any differentiable function $g$,
    \[
        \partial_t g \bpar{D_f(\mu_t\mmid \nu_t)} = -\frac{1}{2}\, g'\bpar{D_f(\mu_t \mmid \nu_t)} \times \E_{\mu_t} \bigl\langle\nabla \bpar{f' \circ \frac{\mu_t}{\nu_t}}, \nabla \log \frac{\mu_t}{\nu_t}\bigr\rangle \,.
    \]
\end{lemma}

\begin{proof}
    The case where $g\neq\msf{id}$ is an application
of the chain rule, so it suffices to take $g=\msf{id}$ and simply
differentiate an $f$-divergence.

For brevity, we drop the variable $x$ from functions involved and proceed by differentiating under the integral sign and discarding boundary terms in integration by parts:
\begin{align*}
\partial_{t}D_{f}(\mu_{t}\mmid\nu_{t}) 
 & =\int\Bbrace{\bpar{f\circ\frac{\mu_{t}}{\nu_{t}}}\,\partial_{t}\nu_{t} + \bpar{f'\circ\frac{\mu_{t}}{\nu_{t}}}\bpar{\frac{\mu_{t}}{\nu_{t}}}'\nu_{t}}\,\D x\\
 & =\int\Bbrace{\partial_{t}\nu_{t}\Bpar{\bpar{f\circ\frac{\mu_{t}}{\nu_{t}}}-\bpar{f'\circ\frac{\mu_{t}}{\nu_{t}}}\frac{\mu_{t}}{\nu_{t}}}+\bpar{f'\circ\frac{\mu_{t}}{\nu_{t}}}\,\partial_{t}\mu_{t}}\,\D x\\
 & \underset{(i)}{=}\int\bbrack{-\nabla\cdot(b_{t}\nu_{t})+\frac{1}{2}\,\Delta\nu_{t}}\Bpar{\bpar{f\circ\frac{\mu_{t}}{\nu_{t}}}-\bpar{f'\circ\frac{\mu_{t}}{\nu_{t}}}\frac{\mu_{t}}{\nu_{t}}}\,\D x\\
 & \qquad+\int\bbrack{-\nabla\cdot(b_{t}\mu_{t})+\frac{1}{2}\,\Delta\mu_{t}}\bpar{f'\circ\frac{\mu_{t}}{\nu_{t}}}\,\D x\,,
\end{align*}
where in $(i)$ we substitute the F-P equation from \eqref{eq:arbitrary_diffusion}.
Integrating by parts (i.e., $\int f\Div(\mb G)=-\int\inner{\nabla f,\mb G}$
for a real-valued function $f$ and vector-valued function $\mb G$),
we have that 
\begin{equation}\label{eq:div_part1}
\int\bbrack{-\nabla\cdot(b_{t}\nu_{t})}\bpar{f\circ\frac{\mu_{t}}{\nu_{t}}}\,\D x=\int\binner{b_{t}\nu_{t},\bpar{f'\circ\frac{\mu_{t}}{\nu_{t}}}\nabla\frac{\mu_{t}}{\nu_{t}}}\,\D x\,.
\end{equation}
On the other hand, we have that 
\[
-\int\bbrack{-\nabla\cdot(b_{t}\nu_{t})}\bpar{f'\circ\frac{\mu_{t}}{\nu_{t}}}\frac{\mu_{t}}{\nu_{t}}\,\D x=-\int\binner{b_{t}\nu_{t},\frac{\mu_{t}}{\nu_{t}}\,\nabla\bpar{f'\circ\frac{\mu_{t}}{\nu_{t}}}+\bpar{f'\circ\frac{\mu_{t}}{\nu_{t}}}\nabla\frac{\mu_{t}}{\nu_{t}}}\,\D x\,.
\]
The second term cancels with the RHS of~\eqref{eq:div_part1}. We have a similar
cancellation for the $\frac{1}{2}\,\Delta\nu_{t}$ term: 
\[
\int\frac{1}{2}\,\Delta\nu_{t}\,\bpar{f\circ\frac{\mu_{t}}{\nu_{t}}}\,\D x=-\int\frac{1}{2}\,\binner{\nabla\nu_{t},\bpar{f'\circ\frac{\mu_{t}}{\nu_{t}}}\nabla\frac{\mu_{t}}{\nu_{t}}}\,\D x\,,
\]
and 
\[
-\int\frac{1}{2}\,\Delta\nu_{t}\,\bpar{f'\circ\frac{\mu_{t}}{\nu_{t}}}\frac{\mu_{t}}{\nu_{t}}\,\D x=\int\frac{1}{2}\,\binner{\nabla\nu_{t},\frac{\mu_{t}}{\nu_{t}}\,\nabla\bpar{f'\circ\frac{\mu_{t}}{\nu_{t}}} + \bpar{f'\circ\frac{\mu_{t}}{\nu_{t}}}\nabla\frac{\mu_{t}}{\nu_{t}}}\,\D x\,.
\]
Combining these, we are left with 
\begin{align*}
\int\bbrack{-\nabla\cdot(b_{t}\nu_{t}) + \frac{1}{2}\,\Delta\nu_{t}}\Bpar{\bpar{f\circ\frac{\mu_{t}}{\nu_{t}}}-\bpar{f'\circ\frac{\mu_{t}}{\nu_{t}}}\frac{\mu_{t}}{\nu_{t}}}\,\D x 
&=-\int\binner{b_{t}\nu_{t}-\frac{1}{2}\,\nabla\nu_{t},\nabla\bpar{f'\circ\frac{\mu_{t}}{\nu_{t}}}\frac{\mu_{t}}{\nu_{t}}}\,\D x\\
 & =-\int\binner{b_{t}\mu_{t}-\frac{1}{2}\,\mu_{t}\nabla\log\nu_{t},\nabla\bpar{f'\circ\frac{\mu_{t}}{\nu_{t}}}}\,\D x\,.
\end{align*}
Finally, we note that 
\begin{align*}
\int\bbrack{-\nabla\cdot(b_{t}\mu_{t})+\frac{1}{2}\Delta\mu_{t}}\bpar{f'\circ\frac{\mu_{t}}{\nu_{t}}}\,\D x & =\int\binner{b_{t}\mu_{t}-\frac{1}{2}\,\nabla\mu_{t},\nabla\bpar{f'\circ\frac{\mu_{t}}{\nu_{t}}}}\,\D x\\
 & =\int\binner{b_{t}\mu_{t}-\frac{1}{2}\,\mu_{t}\nabla\log\mu_{t},\nabla\bpar{f'\circ\frac{\mu_{t}}{\nu_{t}}}}\,\D x\,.
\end{align*}
Putting it all together, noticing that the drift terms cancel, we
are left with 
\[
\partial_{t}D_{f}(\mu_{t}\mmid\nu_{t})=-\int\frac{1}{2}\,\binner{\mu_{t}\nabla\log\frac{\mu_{t}}{\nu_{t}},\nabla\bpar{f'\circ\frac{\mu_{t}}{\nu_{t}}}}\,\D x=-\frac{1}{2}\,\E_{\mu_{t}}\binner{\nabla\log\frac{\mu_{t}}{\nu_{t}},\nabla\bpar{f'\circ\frac{\mu_{t}}{\nu_{t}}}}\,,
\]
which completes the proof.
\end{proof}

To recover the decay result for the $q$-R\'enyi divergence, one can substitute $g(x) = \frac{1}{q-1}\log x$ and $f(x) = x^q - 1$. For the $\chi^2$-divergence, instead substitute $g(x) = x$ and $f(x) = x^2 - 1$. From this, we can obtain a single step of decay for the R\'enyi and $\chi^2$-divergences under different functional inequalities.

\section{Bias from failure}\label{app:bias-from-failure}

By computing the one-step distribution, we can note that $\fb$ conditioned on success has a bias. To see this, consider $\fb$ with threshold $N$ and the initial measure $X_{1}\sim\mu$. Then, for a measurable set $A$ and the success event $S$,
\begin{align*}
\P(X_{2}\in A\mid S) &= 
\frac{\P(X_{2}\in A,S)}{\P(S)}=\frac{1}{\P(S)}\int\P(X_{2}\in A,X_{1}=x,S)\,\D x\\
	&=\frac{1}{\P(S)}\int\P(X_{2}\in A,S\mid X_{1}=x)\,\mu(\D x)\\
	&=\frac{1}{\P(S)}\int\P(X_{2}\in A,S,Y_{2}=y\mid X_{1}=x)\,\mu(\D x)\,\D y\\
	&=\frac{1}{\P(S)}\int\P(X_{2}\in A,S\mid X_{1}=x,Y_{2}=y)\,\gamma(y-x)\,\mu(x)\,\D x\D y\\
	&=\frac{1}{\P(S)}\int\P(X_{2}\in A,S\mid Y_{2}=y)\,\gamma(y-x)\,\mu(x)\,\D x\D y\\
	&=\frac{1}{\P(S)}\int\P(S\mid Y_{2}=y)\P(X_{2}\in A\mid S,Y_{2}=y)\,\gamma(y-x)\,\mu(x)\,\D x\D y\\
	&=\frac{1}{\P(S)}\int\bpar{1-\bpar{1-\ell(y)}^{N}}\,\pi^{X|Y=y}(A)\,\gamma(y-x)\,\mu(x)\,\D x\D y\\
	&=\frac{\int\bpar{1-\bpar{1-\ell(y)}^{N}}\,\pi^{X|Y=y}(A)\,\gamma(y-x)\,\mu(x)\,\D x\D y}{1-\E_{\mu*\gamma}[(1-\ell)^{N}]}\,.
\end{align*}
On the other hand, $\fb$ without cap (i.e., $N=\infty$) satisfies 
\[
\P(X_{2}\in A) =\int\pi^{X|Y=y}(A)\,\gamma(y-x)\,\mu(x)\,\D x\D y\,.
\]
Thus, trajectories are weighted differently in the two distributions. 

\end{document}